\documentclass{article}

\usepackage{microtype}
\usepackage{graphicx}
\usepackage{subcaption}
\usepackage{booktabs} 

\usepackage{hyperref}

\usepackage[accepted]{icml2026}

\usepackage{amsmath}
\usepackage{amssymb}
\usepackage{mathtools}
\usepackage{amsthm}

\usepackage{pifont}
\usepackage{tikz}
\usepackage{adjustbox} % 导入 adjustbox 宏包
\usepackage[table]{xcolor}
\usepackage{filecontents}
\usepackage{booktabs}       % professional-quality tables
\usepackage{amsfonts}       % blackboard math symbols
\usepackage{nicefrac}       % compact symbols for 1/2, etc.
\usepackage{microtype}      % microtypography

\usepackage{microtype}
\usepackage{bbm}
\usepackage{graphicx}
\usepackage{ragged2e}
\usepackage{booktabs,makecell, multirow,tabularx}

\usepackage{algorithm}
\usepackage{algorithmic}
\renewcommand{\algorithmicrequire}{ \textbf{Input:}}
\renewcommand{\algorithmicensure}{ \textbf{Output:}}
\newcommand{\RETURN}{\textbf{return }}

\usepackage{mathtools}
\usepackage{amsthm}
\usepackage{marvosym} 

\usepackage{xcolor}   % for colors
\usepackage[capitalize,noabbrev]{cleveref}

\theoremstyle{plain}

\theoremstyle{definition}

\theoremstyle{remark}

\usepackage{diagbox}

\usepackage[textsize=tiny]{todonotes}
\usepackage[most]{tcolorbox}
\tcbset{
  insightbox/.style={
    colback=gray!5,
    colframe=gray!40,
    boxrule=1pt,
    left=1pt,right=0pt,top=0pt,bottom=2pt,
  }
}
\newcommand{\relDec}[1]{\textcolor{green!50!black}{\scriptsize(\downarrow #1\%)}}

\usepackage{xspace}
\newcommand{\attackname}{\texttt{CAGE}\xspace} 

\usepackage{enumitem}
\usepackage[textsize=tiny]{todonotes}
\icmltitlerunning{On the Adversarial Robustness of Large Vision-Language Models under Visual Token Compression}

\begin{document}

\twocolumn[
  \icmltitle{On the Adversarial Robustness of Large Vision-Language Models \\ under Visual Token Compression}

  % Optional: equal contribution symbol (remove if not needed)

  \begin{icmlauthorlist}
    \icmlauthor{Xinwei Zhang}{polyu}  
    \icmlauthor{Hangcheng Liu}{ntu}
    \icmlauthor{Li Bai}{polyu}
    \icmlauthor{Hao Wang}{cqu}
    \icmlauthor{Qingqing Ye}{polyu}
    \icmlauthor{Tianwei Zhang}{ntu}
    \icmlauthor{Haibo Hu}{polyu,center}
  \end{icmlauthorlist}

  \icmlaffiliation{polyu}{The Hong Kong Polytechnic University, Hong Kong}
  \icmlaffiliation{ntu}{Nanyang Technological University, Singapore}
  \icmlaffiliation{cqu}{Chongqing University, Chongqing, China}
  \icmlaffiliation{center}{Research Centre for Privacy and Security Technologies in Future Smart Systems, PolyU}
  
  % Corresponding authors (edit emails accordingly)
  \icmlcorrespondingauthor{Li Bai}{baili.bai@connect.polyu.hk}

  \icmlkeywords{Machine Learning, ICML}

  \vskip 0.3in
]



\printAffiliationsAndNotice{}  % no special notice (required even if empty)

\begin{abstract}
Visual token compression is widely used to accelerate large vision-language models (LVLMs) by pruning or merging visual tokens, yet its adversarial robustness remains unexplored. We show that existing encoder-based attacks cannot fully disclose the robustness vulnerabilities of compressed LVLMs, due to an optimization-inference mismatch: perturbations are optimized on the full-token representation, while inference is performed through a token-compression bottleneck. To address this gap, we propose the \textbf{C}ompression-\textbf{A}li\textbf{G}n\textbf{E}d attack (\attackname), which aligns perturbation optimization with compression inference without assuming access to the deployed compression mechanism or its token budget. \attackname combines (i) \emph{expected feature disruption}, which concentrates distortion on tokens likely to survive across plausible budgets, and (ii) \emph{rank distortion alignment}, which actively aligns token distortions with rank scores to promote the retention of highly distorted evidence. Across diverse representative plug-and-play compression mechanisms and datasets, our results show that \attackname consistently achieves lower robust accuracy than the baseline. This work highlights that robustness assessments ignoring compression can be overly optimistic, calling for compression-aware security evaluation and defenses for efficient LVLMs.
\end{abstract}

%Our findings highlight the intellectual property risk of LVLMs, calling for increased awareness among developers and the community regarding this risk.

% % \sen{Although model extraction attacks (MEAs) that infringe on model confidentiality via black-box access pose significant risks, their impact on LVLMs remains under-explored.}
% To fill this gap, this paper presents the first study to investigate the feasibility of MEAs against LVLMs. Specifically, we propose \textbf{LVLM-Thief}, a novel query- and resource-efficient framework, that achieves attack by replicating the target model's image-text alignment.
% In particular, to improve query efficiency, we first propose a Robust Consistency Sampling (RCS) method to select robust image samples with high consistency quality.
% Then, we propose a novel Consistency-Guided Augmented Fine-tuning (CGAF) method to explore the complex many-to-many multimodal alignment in LVLMs. By only fine-tuning the modality adapter, the core component responsible for alignment, our attack can also achieve resource efficiency.     
\section{Introduction}
\label{sec:introduction}
The rapid advancement of large vision-language models (LVLMs) has revolutionized multimodal understanding, enabling remarkable capabilities in tasks ranging from visual question answering to complex reasoning \cite{llava,gpt4,gemini}. However, this progress comes with a substantial computational burden: current state-of-the-art models like LLaVA-NeXT~\cite{li2024llavanext} and InternVL~\cite{chen2024internvl} process hundreds to thousands of visual tokens per image, creating significant efficiency bottlenecks that limit their practical deployment. This bottleneck is further amplified in agentic settings (especially mobile agents), where the model must process multiple images under tight latency and energy budgets \cite{zhang2023appagent,ye2025mobile}.

\begin{figure}
    \centering
\includegraphics[width=1\linewidth]{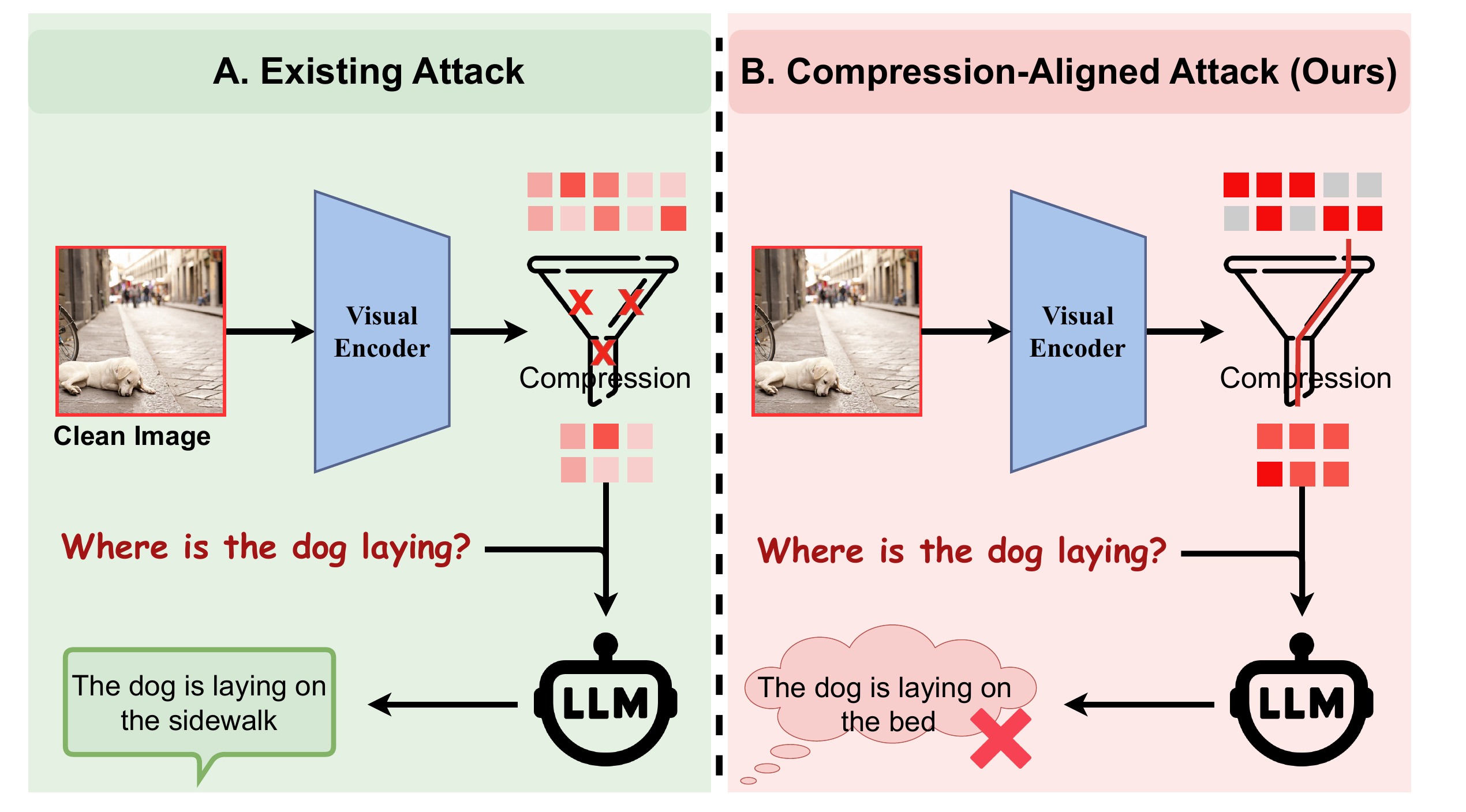}
    \caption{Comparison between the existing attack and our attack. {\color{red}Darker red} indicates tokens with stronger adversarial perturbation. While the existing attack (A) perturbs all visual tokens (all tokens are red), \attackname (B) concentrates the distortion on the surviving tokens (only survivors are red).}
    \label{fig:compare}
    % \vspace{-7mm}
\end{figure}
Given these efficiency constraints, recent research has focused intensively on visual token compression~\cite{liu2024mustdrop,shao2025holitom,bolya2023tome,alvar2025divprune,tongFlowCut2025}. In particular, plug-and-play methods such as VisionZip~\cite{visionzip} and VisPruner~\cite{zhang2025vispruner} identify a compact set of ``informative'' tokens using importance scores (e.g., attention-based saliency), and optionally merge additional tokens via similarity-based merging before passing the compressed visual sequence to the language model. By reducing the visual-token budget at inference time, these approaches deliver substantial efficiency gains while largely preserving accuracy on normal tasks. However, the adversarial robustness of these compressed LVLMs remains unexplored.

Assessing the adversarial robustness of such models is crucial, as token compression has become indispensable for deploying LVLMs in real-time, safety-critical applications (e.g., autonomous driving and robotics) due to strict latency and energy constraints \cite{lubberstedt2025v3lma,survey_vla4ad,guerrero2025efficient}. Existing evaluations typically consider encoder-based attacks, which optimize perturbations over global visual tokens \cite{mei2025veattack, wangBreakVisualPerception2024}. This choice is pragmatic as encoder-only optimization is far cheaper than end-to-end attacks, and vision encoders are often publicly available even when the full LVLM is not~\cite{mei2025veattack, cuiRobustnessLargeMultimodal2024, wangBreakVisualPerception2024}.
However, as illustrated in Figure~\ref{fig:compare}, existing attacks cannot fully disclose the robustness vulnerabilities of compressed LVLMs due to an \textit{optimization-inference mismatch}: perturbations are optimized in the full-token space, whereas model inference operates on the compressed token representation.

To address this mismatch, we propose the \textbf{C}ompression-\textbf{A}li\textbf{G}n\textbf{E}d attack (\attackname), an adversarial framework that aligns perturbation optimization with the compressed token space without assuming knowledge of the deployed compression mechanism or configuration (e.g., token budget). 
The core challenge lies in the uncertainty of deployment budgets, which renders the effective attack feature space unpredictable during optimization. To overcome it, \attackname couples two complementary objectives: (i) \emph{Expected Feature Disruption} (EFD), which models the unknown token budget probabilistically and concentrates distortion on tokens likely to survive the bottleneck; and (ii) \emph{Rank Distortion Alignment} (RDA), which aligns the ranking signal (e.g., attention scores) with distortion strength, so that the most perturbed tokens also have the highest probability
of being selected. 

We evaluate \attackname across five representative compression mechanisms and three datasets. Our results show that \attackname consistently yields substantially lower adversarial accuracy than the existing attack. 
In addition, we investigate some possible defenses. While both defenses show somewhat encouraging results, they are not yet sufficient, highlighting the need for more effective defenses.

% we investigate two mitigation directions: (i) selection-based defenses, and (ii) attention-based detection
% \noindent\textbf{Contributions.} Our main contributions are summarized as follows:
% \ding{172}~We are the \textit{first} to explore the adversarial robustness of LVLMs under visual token compression.
% \ding{173}~We identify an optimization-inference mismatch between existing attacks and compressed inference, which leads to an overestimation of robustness.
% \ding{174}~We propose a novel attack, named \attackname, that remains effective under unknown compression mechanisms and varying deployment token budgets.
% \ding{175}~We demonstrate the effectiveness of our attack across diverse compression mechanisms, token budgets, and datasets, and further explore potential defense strategies.

In summary, the contributions of this paper are as follows.
\begin{itemize}[leftmargin=*, itemsep=0pt, topsep=0pt, parsep=0pt]
\item We are the \textit{first} to explore the adversarial robustness of LVLMs under visual token compression.
\item We identify an optimization-inference mismatch between existing attacks and compressed inference, which leads to an overestimation of robustness.
\item We propose a novel attack, named \attackname\footnote{Source code is available at \url{https://github.com/XinweiZhang1998/CAGE}.}, that remains effective under unknown compression mechanisms and varying deployment token budgets.
\item We demonstrate the effectiveness of our attack across diverse compression mechanisms, token budgets, and datasets, and further explore potential defense strategies. 
\end{itemize}

% \textbf{Concurrent submission.} A concurrent ICML 2026 submission~\cite{ConcurrentSafetyCircuit2026} also considers adversarial vulnerabilities of LVLMs, but targets transferable jailbreak via safety-circuit intervention, whereas we study robustness under visual token compression.

% \noindent\textbf{Contributions.} 
% \ding{172} We are the first to explore the adversarial robustness of compressed LVLMs.
% \ding{173} We identify and empirically validate a structural optimization misalignment between standard global attacks and compressed inference, explaining why robustness can be overestimated under token bottlenecks.
% \ding{174} We introduce \attackname, a compression-aware gray-box attack that remains effective under unknown and varying token budgets via EFD and RDA.
% \ding{175} We conduct extensive evaluations across multiple compression methods, budgets, and benchmarks, and investigate practical defense/detection directions for efficient LVLM deployment.
\section{Preliminaries}
\label{sec:preliminaries}
\subsection{$K$-Compressed LVLM}
A standard LVLM typically consists of a visual encoder $\mathcal{E}$, a visual projector $\mathcal{P}$, and a large language model (LLM) $\mathcal{F}$. Given an input image $\mathbf{x}_v \in \mathbb{R}^{H \times W \times C}$ and a textual prompt $\mathbf{x}_t$ (e.g., a question in VQA tasks), the visual encoder extracts a sequence of visual tokens $\mathbf{H} = \mathcal{E}(\mathbf{x}_v) \in \mathbb{R}^{N \times D}$, where $N$ is the number of tokens (typically $N=576$ or higher) and $D$ is the embedding dimension.

In a $K$-compressed LVLM, a compression module $\mathcal{C}(\cdot; K)$ is introduced to reduce the visual sequence length from $N$ to a smaller budget $K$ ($K < N$) before feeding them into the LLM. In the context of model inference, we refer to this constraint $K$ as the \textbf{deployment budget} (denoted as $K_\text{model}$). This process can be formulated as:
\begin{equation}
    \mathbf{Z} = \mathcal{C}(\mathbf{H}; K) \in \mathbb{R}^{K \times D},
\end{equation}
where $\mathcal{C}$ represents the compression strategy (e.g., token selection/pruning or merging based on attention scores). The compressed visual tokens $\mathbf{Z}$ are then concatenated with the text embeddings of $\mathbf{x}_t$ and fed into the LLM to generate the response $\mathbf{Y}$. The probability of generating the response is modeled auto-regressively:
\begin{equation}
    P(\mathbf{Y} | \mathbf{x}_v, \mathbf{x}_t) = \prod_{i=1}^{L} P(y_i | \mathbf{Z}, \mathbf{x}_t, y_{<i}),
\end{equation}
where $y_{<i}$ denotes the tokens generated before step $i$.

\subsection{Threat Model}
\label{sec:threat_model}

In this paper, we focus on the adversarial robustness of LVLMs under a realistic gray-box threat model, following current attacks \cite{wangBreakVisualPerception2024,mei2025veattack}. 

% \noindent\textbf{Knowledge Assumption.}
% \xw{Some illustration} We assume white-box access to the visual encoder $\mathcal{E}$, as such backbones are widely open-sourced. Crucially, we assume black-box access to the compression module $\mathcal{C}$ and the downstream LLM $\mathcal{F}$, treating the specific compression hyperparameters (e.g., deployment budget $K$) as unknown deployment configurations.

\noindent\textbf{Knowledge Assumption.}
We assume that the adversary has white-box access to the visual encoder $\mathcal{E}$, as such backbones are often open-sourced in practice \cite{wang2024instruct,wangBreakVisualPerception2024,mei2025veattack}. 
In contrast, the adversary has black-box access to both the compression module $\mathcal{C}$ and downstream LLM $\mathcal{F}$. Specifically, we treat compression hyperparameters (e.g., the deployment budget $K_\text{model}$) as unknown runtime variables. This reflects common deployments where token compression is diverse and often customized by the model owner, and the downstream LLM is typically proprietary and not publicly available.

\noindent\textbf{Attack Objective.}
Given an image $\mathbf{x}_v$, question $\mathbf{x}_t$, and ground-truth $\mathbf{Y}_{gt}$, the adversary seeks an imperceptible perturbation $\boldsymbol{\delta}$ to mislead the prediction.
Ideally, the goal is to maximize the negative log-likelihood of the ground-truth answer:
\begin{equation}
    \label{eq:ideal_loss}
    \mathcal{L}(\boldsymbol{\delta}) = - \log P(\mathbf{Y}_{gt} | \mathcal{C}(\mathcal{E}(\mathbf{x}_v + \boldsymbol{\delta}); K), \mathbf{x}_t).
\end{equation}

Under our gray-box setting, backpropagating through the full surrogate LVLM is computationally expensive and prone to overfitting. Therefore, following prior work (e.g., VEAttack~\cite{mei2025veattack}), we adopt an encoder-based attack formulation.
Instead of optimizing the output-text probability, the adversary directly perturbs the vision encoder’s representations under white-box access, making the attack task- and question-agnostic. Specifically, we maximize the semantic deviation by maximizing the cosine distance between clean features $\mathbf{H} = \mathcal{E}(\mathbf{x}_v)$ and adversarial features $\mathbf{H}' = \mathcal{E}(\mathbf{x}_v + \boldsymbol{\delta})$:
\begin{equation}
    \label{eq:enc_objective}
    % \max_{\boldsymbol{\delta}} \mathcal{J}(\boldsymbol{\delta}) = 
    \max_{\boldsymbol{\delta}} \left( 1 - \mathcal{S}(\mathbf{H},\mathbf{H}')\right),
\end{equation}
where $\mathcal{S}(\mathbf{u}, \mathbf{v}) = \frac{\mathbf{u}^\top \mathbf{v}}{\|\mathbf{u}\|\|\mathbf{v}\|}$ denotes the cosine similarity between two vectors. Since we perturb only the visual input, we will denote $\mathbf{x}_v$ simply as $\mathbf{x}$ in the following for notational simplicity.
By pushing adversarial representations away from the clean feature in the visual embedding space, the attack induces erroneous downstream responses without requiring access to the specific prompt.

\begin{figure}
    \centering
    \includegraphics[width=0.6\linewidth]{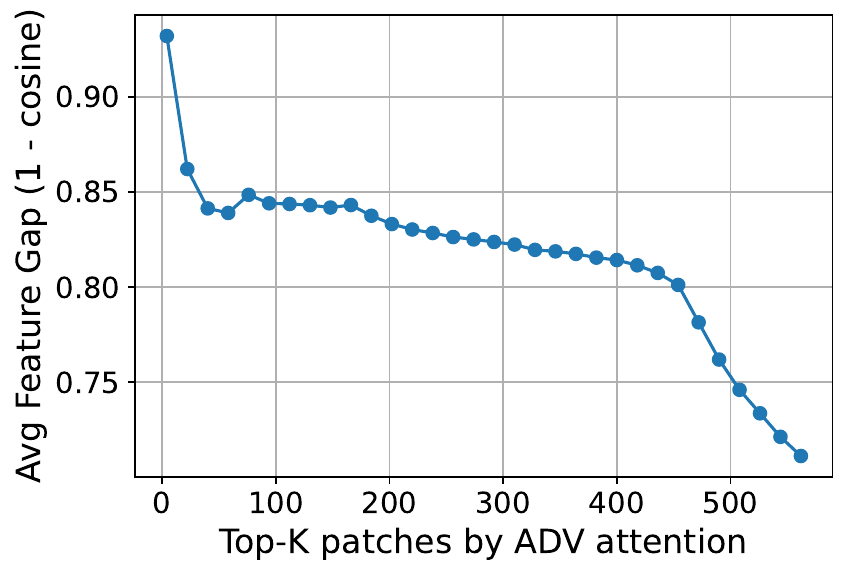}
\caption{Average token-level feature gap under VEAttack over 100 samples.
We rank vision tokens by adversarial(ADV) attention and plot the average feature gap ($1-\text{cosine}$) over the top-$K$ tokens. 
The curve shows that the gap is large on a small number of high-attention tokens and gradually decreases as lower-ranked tokens are included.}
\label{fig:feature_shift}
    % \vspace{-5mm}
\end{figure}

\section{Motivation: An Empirical Study}
\label{sec:motivation}

To understand how visual token compression impacts the adversarial robustness of LVLMs, we conduct an empirical study by applying VEAttack~\cite{mei2025veattack} to LLaVA equipped with VisionZip~\cite{visionzip}. VisionZip reduces the visual token budget via Top-$K$ token selection followed by token aggregation, a common design pattern in recent compressed LVLMs. Our analysis yields two key insights as follows.

% We evaluate multiple token budgets and summarize the results in Table~\ref{tab:pilot_study}.

% As expected, clean accuracy drops noticeably as the token budget decreases. 
% Interestingly, compared to the full model, compressed LVLMs exhibit a noticeably smaller attack-induced accuracy drop under the same evaluation protocol (e.g., $\Delta{=}13.5\text{--}17.7$ pp for compressed models vs.\ $18.7$ pp for the full model).
% \textbf{At face value, this creates an ``illusion of robustness,''} suggesting that aggressively pruned models are inherently more secure. 
% However, we argue that this apparent stability does not reflect intrinsic robustness; instead, it indicates that standard attacks fail to effectively penetrate the compression bottleneck and thus overestimate the robustness of compressed LVLMs.
% Our analysis reveals two critical insights that substantiate this \emph{false sense of security}.

We first investigate whether token pruning inadvertently mitigates adversarial effects by filtering out perturbed tokens.
% Intuitively, one might expect pruning to inadvertently filter out adversarial noise. However, a closer examination of the selection dynamics reveals the opposite. 
\Cref{fig:feature_shift} probes the token-level representation drift by ranking visual tokens according to adversarial attention scores and computing the average feature gap (cosine distance) between clean and adversarial features over the top-$K$ tokens. 
We observe that the average feature gap is larger for smaller $K$ and monotonically decreases as $K$ increases. 
This implies that adversarial distortion is concentrated on a small set of high-importance tokens and is diluted as lower-ranked unimportant tokens are included. Since the compression mechanism preserves high-importance tokens, it keeps the most distorted tokens while discarding weakly perturbed low-importance ones, yielding a ``distortion-concentrated'' survivor set. Consequently, the compressed model bases its prediction on the most strongly corrupted visual evidence rather than a denoised representation.
\begin{tcolorbox}[insightbox]
\textit{\textbf{Insight I}: Under adversarial inputs, compression behaves as a ``distortion concentrator'' rather than a denoiser: it preferentially retains heavily corrupted, high-importance tokens while discarding relatively clean, low-importance tokens.}
\end{tcolorbox}

\begin{table}[t]
    \centering
    \caption{Robust accuracy (\%, $\downarrow$) on compressed LLaVA when varying the model’s token budget $K_{\text{model}}$ (rows) and the attack optimization budget $K_{\text{attack}}$ (columns).
    For each deployment setting, the most harmful configuration (in \textbf{bold}) typically corresponds to an attack budget aligned with the model’s effective token budget, rather than the full-token baseline.}
    \label{tab:budget_mismatch}
    \resizebox{0.7\linewidth}{!}{
    \begin{tabular}{c|c|ccc}
        \toprule
        \diagbox[width=6em]{\textbf{$K_{\text{model}}$}}{\textbf{$K_{\text{attack}}$}}
        & \textbf{576 (Full)} & \textbf{192} & \textbf{64} & \textbf{16} \\
        \midrule
        % \textbf{576 Tokens (Full)}       & 55.8 & \textbf{51.0} & 56.0 & 61.4  \\
        \textbf{192} & 55.7  & \textbf{50.7} & 53.2  & 60.4  \\
        \textbf{64}  & 53.5  & 48.8  & \textbf{48.7}  & 56.0  \\
        \textbf{16}  & 49.7  & 45.3  & 44.7  & \textbf{44.4}  \\
        \bottomrule
    \end{tabular}
    }
\end{table}

While Insight~I shows that compression selects heavily distorted survivor tokens, we next investigate whether existing attacks remain maximally harmful under such compressed inference.
% To probe this, we introduce an attack optimization budget $K_{\text{attack}}$ and modify the VEAttack objective by computing the loss only on the Top-$K_{\text{attack}}$ visual tokens ranked by attention (computed under the current forward pass).
% Importantly, the perturbation is still applied to the entire input image; $K_{\text{attack}}$ only controls which tokens contribute to the optimization signal.
To probe this, we introduce an attack optimization budget $K_{\text{attack}}$ and modify the VEAttack objective to compute the loss only on the Top-$K_{\text{attack}}$ visual tokens.
Let $\text{TOP}_{K_{\text{attack}}}(\cdot)$ denote the subset of $K_{\text{attack}}$ token features with the largest attention scores, where the attention scores are computed on the current adversarial forward pass and updated at each optimization step. We optimize
\begin{equation}
\label{eq:topk_attack_obj}
\max_{\boldsymbol{\delta}} \Big( 1 - \mathcal{S}\big(\text{TOP}_{K_{\text{attack}}}(\mathbf{H}),\ \text{TOP}_{K_{\text{attack}}}(\mathbf{H}')\big) \Big).
\end{equation}
% Importantly, $\boldsymbol{\delta}$ is still applied to the entire input image; $K_{\text{attack}}$ only determines which tokens contribute gradients during optimization.
As shown in Table~\ref{tab:budget_mismatch}, the standard full-token setting ($K_{\text{attack}}{=}\text{Full}$) tends to be overly optimistic under compressed inference: restricting optimization to a compression-aware token budget yields lower robust accuracy.
This indicates a clear \textbf{optimization-inference mismatch} in existing attacks: perturbations are optimized over the full token space, whereas inference depends on a compressed token representation.
% This reveals a first-level \textbf{attack--compression mismatch}: standard attacks are optimized in the pre-compression full-token representation, while deployment-time inference depends on the post-compression tokens. As a result, a large fraction of the optimization effort does not translate to the compressed inference pathway, leading to overly optimistic robustness estimates.
Under compression, existing attacks that optimized on full token space can become suboptimal due to:
(i) \textit{Budget Dilution}: a non-trivial portion of the optimization signal is allocated to tokens that are later pruned and therefore never influence inference; and
(ii) \textit{Dependency Disruption}: pruning context/background tokens can alter cross-token interactions, potentially weakening the attack effect that global objectives rely on.

Moreover, attack effectiveness depends on budget alignment: attacks are typically strongest when $K_{\text{attack}}$ is comparable to the deployment budget $K_{\text{model}}$.
For example, on the 16-token model, the full-token attack yields 49.7\% robust accuracy, whereas the aligned setting ($K_{\text{attack}}{=}16$) reduces it to 44.4\%. This motivates us to align attack optimization with the post-compression feature space that the deployed model actually uses.
\begin{tcolorbox}[insightbox]
\textit{\textbf{Insight II}: Token compression creates an optimization-inference mismatch, making existing attacks suboptimal and overly optimistic.}
\end{tcolorbox}

\textbf{In summary}, our analysis highlights a critical gap in robustness evaluation for token-compressed LVLMs.
Compression forces the model to rely on heavily distorted survivor tokens (Insight~I), yet existing attacks optimize for the full-token space, ignoring the post-compression inference pathway.
This optimization-inference mismatch (Insight~II) results in optimistic robustness estimates.
These findings motivate us to propose an attack that aligns perturbation optimization with the token compression mechanism.

\section{Compression-Aligned Attack}
\label{sec:method}
\begin{figure}[t]
    \centering
    \includegraphics[width=\linewidth]{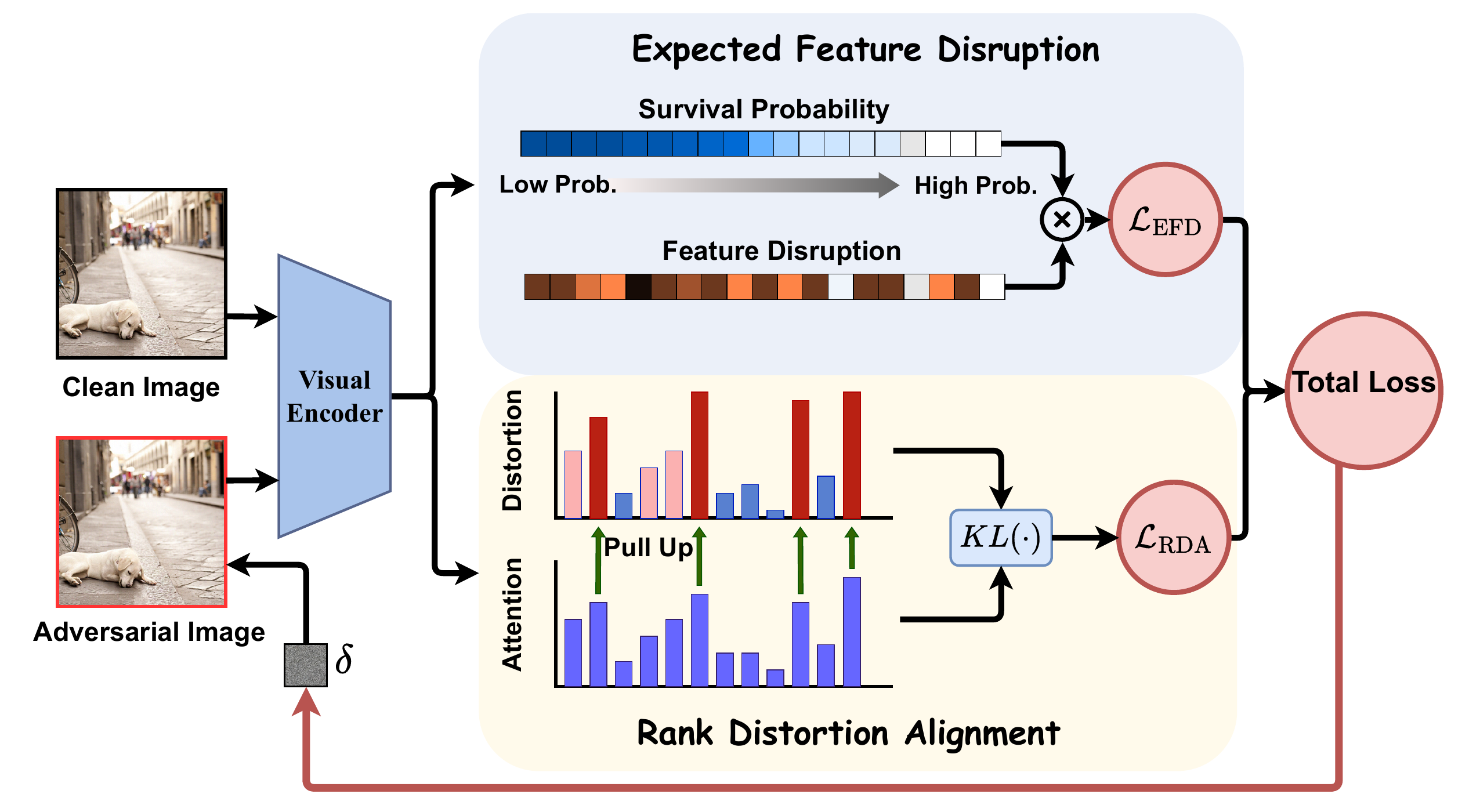} 
    \caption{Overview of \attackname.} 
    % Given a frozen LVLM encoder, \attackname optimizes a perturbation $\boldsymbol{\delta}$ via two synergistic components to overcome the unknown compression bottleneck.
    % \textbf{Top path (EFD):} Models the unknown budget $K$ probabilistically to generate a survival mask $\pi$, concentrating feature distortion $d$ on likely survivors (the "Payload").
    % \textbf{Bottom path (RDA):} Explicitly aligns the model's attention distribution $p_s$ with the distortion distribution $p_d$ via KL-divergence minimization, ensuring corrupted tokens are prioritized for selection (the "Delivery").
    % The joint optimization creates a self-reinforcing loop robust across varying compression budgets.}
    \label{fig:method_overview}
    % \vspace{-8mm}
\end{figure}

Motivated by the findings in \Cref{sec:motivation}, we propose a compression-aligned attack (named \attackname). \attackname bridges the optimization-inference mismatch by reformulating the adversarial objective: instead of indiscriminately disrupting global visual features, \attackname targets the compressed representation that the victim model actually consumes at inference time.
Concretely, \attackname concentrates the perturbation on the limited set of survivor tokens that pass through the token-selection bottleneck, without assuming knowledge of the deployment token budget.

\subsection{Challenge and Overview}
\noindent\textbf{Key challenge.}
An effective attack must concentrate perturbations on the survivor tokens that pass the token-selection bottleneck. However, the deployment token budget $K_{\text{model}}$ is unknown and varies across models.
This uncertainty makes the attack budget $K_{\text{attack}}$ non-trivial: under-estimating $K_{\text{model}}$ leaves part of the survivor set unperturbed (incomplete coverage), whereas over-estimating $K_{\text{model}}$ allocates optimization signal to tokens that will not survive selection (budget dilution), as presented in \Cref{tab:budget_mismatch}.

% because it is a shared first step in common compression select-then-(optional) merge pipelines: 

\noindent\textbf{Overview.}
To address the above challenge, we propose \attackname, a probabilistic attack framework designed to optimize perturbations through an uncertain compression bottleneck.
In \attackname, we strategically target the token selection stage because it is a shared first step in general compression pipelines: some methods perform selection only~\cite{hired2025aaai,alvar2025divprune}, while others first select representative tokens and then merge remaining tokens into them~\cite{visionzip,shang2025prumerge}.
Attacking this selection stage provides a unified interface that naturally covers currently used compression mechanisms.
As shown in \Cref{fig:method_overview}, \attackname mainly includes two components:
\begin{itemize}[leftmargin=*, itemsep=2pt, topsep=2pt, parsep=0pt]
    \item \textit{Expected Feature Disruption (EFD)} maximizes the expected feature disruption that is propagated through compression. It models the deployment budget $K_{\text{model}}$ as a random variable and reweights the distortion of each token by its retention probability, focusing on the tokens likely to survive the bottleneck.
    \item \textit{Rank Distortion Alignment (RDA)} aligns selection scores (e.g., attention) with token distortions via a differentiable distribution-matching loss. This increases the probability that highly distorted tokens are ranked within the selected set and thus influence the compressed representation.
\end{itemize}

\subsection{Expected Feature Disruption}
\label{sec:efd}
A naive way to align the attack with the compression bottleneck is to weight token-level feature disruption by the importance scores $s_i(\mathbf{x})$. However, this is often suboptimal because attention scores are typically highly peaky: most mass concentrates on a few top-ranked tokens even after normalization.
As a result, $s_i(\mathbf{x})$-weighted objectives tend to over-focus on the very top ranks and provide little optimization signal for intermediate tokens.
This is undesirable because a non-trivial portion of these intermediate tokens can still be selected under compression, especially when the deployment budget is moderate or unknown.

To address this, we instead focus on each token's survival probability.
Since the deployment budget $K_{\text{model}}$ is unknown, we model it as a discrete random variable drawn from a prior distribution $P(K_{\text{model}})$ (e.g., $K_{\text{model}}\sim \mathcal{U}[K_{\min}, K_{\max}]$).
Let $r_i(\mathbf{x})$ denote the rank of token $i$ obtained by sorting selection scores $s_i(\mathbf{x})$ in descending order (with rank starting from $0$).
Under Top-$K$ selection, token $i$ is retained if and only if $r_i(\mathbf{x})<K_{\text{model}}$. We thus define its \textit{survival probability} as: 
\begin{equation}
\begin{aligned}
\label{eq:survival_prob}
\pi_i(\mathbf{x})
&\;=\;
P\!\left(K_{\text{model}}>r_i(\mathbf{x})\right) \\
&\;=\;
\sum_{k} P(K_{\text{model}}{=}k)\cdot \mathbb{I}\!\left(r_i(\mathbf{x})<k\right),
\end{aligned}
\end{equation}
where $\mathbb{I}(\cdot)$ denotes the indicator function.
Under a uniform prior, $\pi_i(\mathbf{x})$ forms a rank-decaying soft mask: it equals $1$ for top-ranked tokens ($r_i<K_{\min}$), gradually decays for intermediate ranks, and becomes $0$ for low-ranked tokens ($r_i\ge K_{\max}$).
This allows us to emphasize tokens that are consistently retained across plausible deployment budgets, without committing to a single fixed $K_{\text{model}}$ during attack optimization.

We then use $\pi_i(\mathbf{x})$ to define an objective that maximizes the feature disruption propagated through the compression bottleneck.
Let $\mathbf{z}_i^{\mathrm{cln}}$ and $\mathbf{z}_i^{\mathrm{adv}}$ denote the clean and adversarial feature vectors of token $i$ at the selection layer.
We quantify token-level disruption using cosine distance:
\begin{equation}
d_i(\mathbf{x}) \;=\; 1 - \mathcal{S}\!\left(\mathbf{z}_i^{\mathrm{adv}}, \mathbf{z}_i^{\mathrm{cln}}\right).
\end{equation}
% where $\mathcal{S}(\mathbf{u}, \mathbf{v}) = \frac{\mathbf{u}^\top \mathbf{v}}{\|\mathbf{u}\|\|\mathbf{v}\|}$ denotes the cosine similarity between two vectors.
The EFD loss is computed as the probability-weighted average distortion:
\begin{equation}
\label{eq:efd}
\mathcal{L}_{\text{EFD}}(\mathbf{x}) = 
\frac{\sum_{i=1}^{N} \pi_i(\mathbf{x}) \cdot d_i(\mathbf{x})}{\sum_{i=1}^{N} \pi_i(\mathbf{x})}.
\end{equation}
By weighting distortions with $\pi_i(\mathbf{x})$, $\mathcal{L}_{\text{EFD}}$ concentrates perturbation energy exclusively on high-probability survivors. 
This formulation simultaneously addresses the twin failure modes identified in \Cref{sec:motivation}: 
It mitigates \textit{budget dilution} by ignoring tokens destined for pruning (where $\pi_i \to 0$), and circumvents \textit{dependency disruption} by optimizing the adversarial representation directly within the compressed view, preventing the attack from relying on fragile interactions with vanishing background context.

\subsection{Rank Distortion Alignment}
\label{sec:rda}
While EFD concentrates perturbations on tokens that are likely to survive compression, it does not explicitly align selection with perturbation strength.
Ideally, we want a state of rank-distortion alignment, where the tokens carrying the strongest adversarial perturbations are also assigned the highest selection scores to guarantee their retention.

\noindent\textbf{Why EFD is not enough?}
In theory, EFD would implicitly encourage such alignment. Applying the chain rule to \Cref{eq:efd} yields
\begin{equation}
\label{eq:grad_decomp}
\nabla_{\mathbf{x}}\mathcal{L}_{\text{EFD}}
\;\propto\;
\sum_{i=1}^{N}
\underbrace{\pi_i \cdot \nabla_{\mathbf{x}} d_i}_{\text{survivor corruption}}
\;+\;
\underbrace{d_i \cdot \nabla_{\mathbf{x}}\pi_i}_{\text{distortion promotion}}.
\end{equation}
The first term, $\sum_i \pi_i \nabla_{\mathbf{x}} d_i$, increases distortion on tokens weighted by their current retention probabilities, i.e., it corrupts the tokens that are already likely to be retained.
The second term, $\sum_i d_i \nabla_{\mathbf{x}}\pi_i$, would encourage rank-distortion alignment by pushing the retention weights toward highly distorted tokens, making them more likely to be selected.
However, $\pi_i(\mathbf{x})$ is induced by discrete ranking and Top-$K$ selection, which introduce non-smooth, piecewise-constant dependencies on the underlying scores.
Backpropagating through $\pi_i$ typically yields sparse and unstable gradients (and can be ill-defined at switching points), providing only a limited signal for consistently promoting highly distorted tokens in the ranking.
Consequently, optimization mainly increases distortion on tokens that are already selected/likely retained, but does not reliably steer the selection mechanism toward the most distorted tokens.

% However, $\pi_i(\mathbf{x})$ depends on the discrete ranking operator $r_i(\mathbf{x})$ and Top-$K$ selection, which is non-differentiable in practice; consequently, the ``promotion'' term $d_i\nabla_{\mathbf{x}}\pi_i$ is largely missing or unstable under surrogates. Empirically, this can lead to a failure mode where distortion increases on currently-selected tokens without reliably steering the selection distribution toward the most distorted regions.

To explicitly enforce alignment, we introduce a differentiable \textit{Rank Distortion Alignment (RDA)} objective.
We convert $d_i$ and $s_i^{\mathrm{adv}}$ into token-wise probability distributions via a softmax:
\begin{equation}
p_i^{(d)}(\mathbf{x})=\frac{\exp(d_i(\mathbf{x}))}{\sum_j\exp(d_j(\mathbf{x} ))}, \,
p_i^{(s)}(\mathbf{x})=\frac{\exp(s_i^{\mathrm{adv}}(\mathbf{x}))}{\sum_j\exp(s_j^{\mathrm{adv}}(\mathbf{x}))}.
\end{equation}
We then align the selection mechanism to the distortion profile by maximizing the expected log-likelihood of the selection distribution with respect to the distortion target:
\begin{equation}
\label{eq:rda}
\mathcal{L}_{\text{RDA}}(\mathbf{x})
=
\sum_{i=1}^{N} p_i^{(d)}(\mathbf{x})\,\log p_i^{(s)}(\mathbf{x}).
\end{equation}
During optimization, we treat $p^{(d)}$ as a fixed target and stop gradients through $p^{(d)}$ to avoid degenerate dynamics where both distributions drift simultaneously.
This encourages the selection mechanism to prioritize highly distorted tokens, increasing the chance that adversarial evidence is propagated through the compression bottleneck.

\subsection{Joint Optimization}
\label{sec:opt}

We generate adversarial perturbations $\boldsymbol{\delta}$ using projected gradient descent (PGD) under an $\ell_\infty$ constraint. At each iteration, we compute: (i) the survival probabilities $\pi_i$ from the current adversarial selection scores, (ii) the expected feature disruption $\mathcal{L}_{\text{EFD}}$ via \Cref{eq:efd}, and (iii) the rank-distortion alignment objective $\mathcal{L}_{\text{RDA}}$ via \Cref{eq:rda}. The total optimized objective is given by:
\begin{equation}
\label{eq:total}
\max_{\boldsymbol{\delta}} \;\;
\mathcal{L}_{\text{total}}
\;=\;
\mathcal{L}_{\text{EFD}}
\;+\;
\lambda \cdot \mathcal{L}_{\text{RDA}},
\qquad
\text{s.t. }\|\boldsymbol{\delta}\|_{\infty}\le \epsilon,
\end{equation}
where $\lambda$ is a hyperparameter that balances $\mathcal{L}_{\text{EFD}}$ and $\mathcal{L}_{\text{RDA}}$.
Intuitively, $\mathcal{L}_{\text{EFD}}$ produces high-distortion features on likely survivors (payload), while $\mathcal{L}_{\text{RDA}}$ ensures that the model prioritizes these distorted tokens during selection (delivery), yielding a robust attack on the compressed representation across varying and unknown budgets. The whole procedure is given in \Cref{alg:caa} in Appendix.

% \begin{equation}
% p_d(i)=\frac{\exp(d_i/\tau)}{\sum_j\exp(d_j/\tau)},\qquad
% p_s(i)=\frac{\exp(s_i^{\mathrm{adv}}/\tau)}{\sum_j\exp(s_j^{\mathrm{adv}}/\tau)}.
% \end{equation}
% where $\tau$ is the temperature parameter controlling the sharpness. 
% We apply the temperature only to $p_s$ because the distortion $d_i$ (cosine distance) is naturally bounded and relatively well-scaled, whereas selection scores $s_i^{\mathrm{adv}}$ can be highly peaky and vary in magnitude across layers/models. Thus, $\tau$ mainly serves to calibrate the sharpness of the selection distribution and stabilize the alignment signal, while keeping $p_d$ as a scale-consistent target.
\begin{table*}[t]
\centering
\caption{Main results comparing clean and robust accuracy (\%) across five representative token compression mechanisms on LLaVA. We report \textbf{Clean} accuracy and \textbf{Robust} accuracy under both the baseline attack (\textbf{Base}) and our proposed \attackname. A lower robust accuracy indicates a stronger attack. \textcolor{green!50!black}{Green annotations} denote the reduction in robust accuracy achieved by \attackname compared to the baseline.}
\label{tab:main_result}
\resizebox{\textwidth}{!}{
\begin{tabular}{c|ccl|ccl|ccl}
\toprule
\multirow{2}{*}{\textbf{Compressed Method}} &
\multicolumn{3}{c|}{\textbf{GQA}} &
\multicolumn{3}{c|}{\textbf{TextVQA}} &
\multicolumn{3}{c}{\textbf{VQA-v2}} \\
& Clean &  Robust (Base) & Robust (\attackname)
& Clean & Robust (Base) & Robust (\attackname)
& Clean & Robust (Base) & Robust (\attackname)\\
\midrule

% -------------------------------------------------------
\rowcolor{gray!15}\multicolumn{10}{c}{\textit{Upper Bound (576 Tokens)}} \\
None  & 60.3 & 42.3 & 39.4$_{\relDec{6.9}}$
     & 57.5 & 34.7 & 26.50$_{\relDec{23.6}}$
     & 74.5 & 55.8 & 49.4$_{\relDec{11.4}}$ \\
\midrule

% -------------------------------------------------------
\rowcolor{gray!15}\multicolumn{10}{c}{\textit{Retain 192 Tokens ($K_\text{model}=192$)}} \\
VisionZIP {\scriptsize\textsf{(CVPR25)}}   & 57.0 & 40.9 & 36.2  & 56.9 & 34.1 & 24.6 & 73.4 & 55.7 & 46.5 \\
VisPruner {\scriptsize\textsf{(ICCV25)}}   & 55.3 & 40.8 & 36.5  & 58.2 & 33.9 & 23.0 & 73.4 & 56.3 & 46.4 \\
DivPrune  {\scriptsize\textsf{(CVPR25)}}   & 55.6 & 40.5 & 37.4  & 54.6 & 32.7 & 23.5 & 73.1 & 53.9 & 49.0 \\
FlowCut   {\scriptsize\textsf{(NeurIPS25)}}& 55.4 & 40.3 & 35.1  & 57.8 & 33.6 & 24.3 & 72.9 & 55.9 & 47.3 \\
PruMerge  {\scriptsize\textsf{(ICCV25)}}   & 57.2 & 41.5 & 36.0  & 56.0 & 33.3 & 21.6 & 74.2 & 55.1 & 46.8 \\
\rowcolor{green!03}
\textbf{Average} & 56.1 & 40.8 & 36.2$_{\relDec{11.3}}$
                & 56.7 & 33.5 & 23.4$_{\relDec{30.1}}$
                & 73.4 & 55.4 & 47.2$_{\relDec{14.8}}$ \\
\midrule

% -------------------------------------------------------
\rowcolor{gray!15}\multicolumn{10}{c}{\textit{Retain 128 Tokens ($K_\text{model}=128$)}} \\
VisionZIP {\scriptsize\textsf{(CVPR25)}}   & 55.1 & 40.9 & 35.6  & 57.0 & 33.5 & 21.6 & 72.1 & 53.8 & 45.1 \\
VisPruner {\scriptsize\textsf{(ICCV25)}}   & 54.3 & 41.5 & 36.2  & 56.5 & 34.1 & 21.5 & 72.2 & 56.3 & 46.8 \\
DivPrune  {\scriptsize\textsf{(CVPR25)}}   & 55.6 & 40.4 & 36.5  & 52.7 & 32.2 & 23.2 & 72.2 & 54.0 & 49.2 \\
FlowCut   {\scriptsize\textsf{(NeurIPS25)}}& 54.8 & 40.6 & 34.7  & 57.5 & 33.2 & 21.3 & 70.2 & 54.7 & 44.9 \\
PruMerge  {\scriptsize\textsf{(ICCV25)}}   & 55.5 & 41.0 & 35.3  & 55.1 & 32.9 & 21.9 & 72.5 & 55.8 & 47.8 \\
\rowcolor{green!03}
\textbf{Average} & 55.1 & 40.9 & 35.7$_{\relDec{12.7}}$
                & 55.8 & 33.2 & 21.9$_{\relDec{34.0}}$
                & 71.8 & 54.9 & 46.8$_{\relDec{14.8}}$ \\

\midrule

% -------------------------------------------------------
\rowcolor{gray!15}\multicolumn{10}{c}{\textit{Retain 64 Tokens ($K_\text{model}=64$)}} \\
VisionZIP {\scriptsize\textsf{(CVPR25)}}   & 52.8 & 40.8 & 34.3  & 54.8 & 32.0 & 18.3 & 69.3 & 53.5 & 43.0 \\
VisPruner {\scriptsize\textsf{(ICCV25)}}   & 53.5 & 40.3 & 34.5  & 54.6 & 31.0 & 18.9 & 70.9 & 53.7 & 45.2 \\
DivPrune  {\scriptsize\textsf{(CVPR25)}}   & 52.9 & 40.2 & 36.9  & 50.1 & 29.5 & 21.2 & 71.5 & 54.3 & 48.3 \\
FlowCut   {\scriptsize\textsf{(NeurIPS25)}}& 50.2 & 38.6 & 34.7  & 54.7 & 31.6 & 18.3 & 66.8 & 54.3 & 44.4 \\
PruMerge  {\scriptsize\textsf{(ICCV25)}}   & 53.8 & 41.2 & 34.1  & 52.8 & 31.5 & 17.7 & 70.7 & 54.3 & 43.9 \\
\rowcolor{green!03}
\textbf{Average} & 52.6 & 40.2 & 34.9$_{\relDec{13.2}}$
                & 53.4 & 31.1 & 18.9$_{\relDec{39.2}}$
                & 69.8 & 54.0 & 45.0$_{\relDec{16.7}}$ \\

\midrule

% -------------------------------------------------------

\rowcolor{gray!15}\multicolumn{10}{c}{\textit{Retain 32 Tokens ($K_\text{model}=32$)}} \\
VisionZIP {\scriptsize\textsf{(CVPR25)}}   & 50.1 &40.0  &33.3  & 49.3 & 30.3 & 17.9 &68.4  & 51.3 & 43.1\\
VisPruner {\scriptsize\textsf{(ICCV25)}}   &49.1  & 37.9 & 33.4  & 49.2 & 30.7 & 17.8 &66.9  &51.5  &43.8\\
DivPrune  {\scriptsize\textsf{(CVPR25)}}   & 50.5 & 40.2 & 35.0  & 45.5 &  27.8& 20.3 & 68.7 &52.4
&45.2 \\
FlowCut   {\scriptsize\textsf{(NeurIPS25)}}& 47.2 & 37.2 &33.9   & 50.1 & 29.3 &17.4  &62.2  & 52.6  &44.1\\
PruMerge  {\scriptsize\textsf{(ICCV25)}}   &50.4 & 39.7  & 33.9  &  46.1& 27.6 &  17.5&  67.6& 52.8 &43.6\\
\rowcolor{green!03}
\textbf{Average} 
& 49.5 & 39.0 & 33.8$_{\relDec{13.3}}$
& 48.0 & 29.1 & 18.2$_{\relDec{37.5}}$
& 66.8 & 52.1 & 44.0$_{\relDec{15.5}}$ \\
\midrule

\rowcolor{gray!15}\multicolumn{10}{c}{\textit{Retain 16 Tokens ($K_\text{model}=16$)}} \\
VisionZIP {\scriptsize\textsf{(CVPR25)}}   & 47.4 & 37.2 & 32.4  & 45.8 & 27.2 & 15.8 & 63.2 & 49.7 & 42.7 \\
VisPruner {\scriptsize\textsf{(ICCV25)}}   & 45.9 & 38.9 & 32.8  & 38.2 & 25.1 & 14.7 & 60.4 & 50.9 & 42.4 \\
DivPrune  {\scriptsize\textsf{(CVPR25)}}   & 49.3 & 41.2 & 34.1  & 38.5 & 24.3 & 17.0 & 63.8 & 50.1 & 42.9 \\
FlowCut   {\scriptsize\textsf{(NeurIPS25)}}& 40.7 & 37.8 & 33.6  & 38.7 & 25.7 & 15.1 & 55.5 & 50.3 & 44.6 \\
PruMerge  {\scriptsize\textsf{(ICCV25)}}   & 47.6 & 38.5 & 32.4  & 39.5 & 25.4 & 15.7 & 65.2 & 50.5 & 41.8 \\
\rowcolor{green!03}
\textbf{Average} & 46.2 & 38.7 & 33.1$_{\relDec{14.6}}$
                & 40.1 & 25.5 & 15.7$_{\relDec{38.4}}$
                & 61.6 & 50.3 & 42.9$_{\relDec{14.7}}$ \\
\midrule

\rowcolor{gray!15}\multicolumn{10}{c}{\textit{Blind ($K_\text{model}=0$)}} \\
None  & 21.7 & 21.7 & 21.7 & 10.3 & 10.3 & 10.3 & 44.3 & 44.3 & 44.3 \\
\bottomrule
\end{tabular}}
\end{table*}

\section{Experiments}
\label{sec:experiment}

\subsection{Experimental Setup}
\label{sec:experimental_setup}

\textbf{Victim Models.}
Our evaluation primarily focuses on the widely adopted LLaVA ~\cite{llava}, given its prevalent use as a backbone model in existing token compression research. To evaluate the cross-model generalization of our attack, we also conduct experiments on Qwen2.5-VL~\cite{qwen2025qwen25technicalreport} with the results in Appendix \ref{appendix:qwen}.

\textbf{Compression Mechanisms.}
To ensure broad coverage of compression mechanisms, we evaluate five diverse methods:
VisionZIP \cite{visionzip},
VisPruner \cite{zhang2025vispruner},
DivPrune \cite{alvar2025divprune},
FlowCut \cite{tongFlowCut2025},
and PruMerge \cite{shang2025prumerge}. The detailed descriptions of these mechanisms are given in Appendix \ref{appendix:compression_method}.
We assess these methods across a broad spectrum of deployment token budgets:
(i) \textit{Full Sequence (Upper Bound):} no compression, retaining all $K_\text{model}{=}576$ tokens for LLaVA;
(ii) \textit{Mild to Extreme Compression:} retaining $K_\text{model} \in \{192, 128, 64, 32, 16\}$ tokens; and
(iii) \textit{Blind:} an extreme setting ($K_\text{model}{=}0$) serving as a reference, where no visual tokens are provided to the LLM, so the model responds only based on the language priors.

\textbf{Datasets \& Evaluation Metric.}
We evaluate on three established datasets evaluating diverse capabilities: VQA-v2~\cite{goyal2017making}, TextVQA~\cite{singh2019towards}, and GQA~\cite{hudson2019gqa}. We uniformly sample 1,000 question-image pairs from each dataset for evaluation.
To standardize evaluation across free-form LVLM outputs, we employ a uniform answer-matching protocol: following standard text normalization (lowercasing, removing punctuation/articles, and canonicalizing whitespace), a prediction is deemed correct if it matches any ground-truth answer via a whole-word containment check.
We report \textbf{clean accuracy} and \textbf{robust accuracy}, computed as the percentage of correct predictions on clean and adversarial inputs, respectively.

\textbf{Baseline.}
We adopt VEAttack~\cite{mei2025veattack}, a state-of-the-art encoder-based attack, as our primary baseline. Both VEAttack and our method follow the same gray-box threat model: white-box access to the vision encoder and black-box access to the downstream LLM. This strict alignment ensures that VEAttack serves as a fair and directly comparable baseline in our evaluation.

\textbf{Implementation Setting.}
We generate adversarial perturbations using $\ell_\infty$-bounded PGD with a budget $\epsilon=2/255$, step size $\alpha=0.5/255$, and $T=100$ iterations with random initialization.
Regarding the configuration of \attackname, we utilize the attention weights in the vision encoder as the token importance score $s_i(\mathbf{x})$.
To capture budget uncertainty in the EFD component, we employ a uniform prior $P(K_{\text{model}})$ over the range $[16, 192]$ to cover mild-to-extreme compression.
The balancing hyperparameter is set to $\lambda=0.005$.
All experiments are conducted on NVIDIA RTX 4090 GPUs. 

% Attacks are generated via $\ell_\infty$-bounded PGD with a maximum perturbation budget of $\epsilon=2/255$. The optimization runs for $T=100$ iterations with a step size of $\alpha=0.5/255$ and random initialization within the $\epsilon$-ball.
% For \attackname, we set the hyperparameter $\lambda=0.005$. For the EFD component, we assume a uniform prior distribution $P(K_{\text{model}})$ over the range $[K_{\min}, K_{\max}] = [16, 192]$ (covering the spectrum from mild compression to extreme sequence). We use the attention weight as the importance score $s_i(\mathbf{x})$.
% We conduct all experiments on NVIDIA RTX 4090 GPUs.
% implement all methods using PyTorch and
\subsection{Main Results} 
\label{sec:main_results}

\Cref{tab:main_result} reports attack results against five representative token-compression mechanisms on LLaVA.
Across all settings, \attackname yields consistently lower robust accuracy than the baseline, indicating that compression-aware optimization exposes stronger vulnerabilities under compressed inference.
In addition, our results reveal the following findings.

\textbf{Finding 1: \attackname improves attack efficacy across diverse compression mechanisms, including diversity-based selection.} Diversity-driven methods (e.g., DivPrune) often exhibit higher robust accuracy than attention-based mechanisms, as their retained tokens are less aligned with the attention cues primarily exploited by \attackname. Nevertheless, \attackname still yields consistent gains over standard baselines. The gains persist because attention-salient regions are frequently represented within the diverse token set retained by similarity-based selection. Therefore, \attackname can still inject adversarial evidence into the survivor set even under mismatched selection criteria.

\textbf{Finding 2: \attackname is particularly effective on text-centric visual understanding tasks.}
The relative gain of \attackname is most pronounced on TextVQA, where answers hinge on sparse, highly localized text evidence with limited redundancy, making predictions disproportionately sensitive to the corruption of a few text-bearing survivor tokens under compression.
\attackname is designed to concentrate distortion on likely-retained, high-importance tokens and thus more reliably corrupts these decisive text regions than the baseline.
As a result, the performance drop is amplified on TextVQA relative to VQA-v2 and GQA, where visual reasoning typically aggregates more distributed and redundant cues.

\textbf{Finding 3: \attackname yields consistent advantages across the compression spectrum.}
% , with gains becoming more pronounced under moderate compression and saturating in extreme low-budget regimes
Even without compression (full tokens), \attackname still outperforms the baseline, which we attribute to EFD’s token reweighting that focuses perturbation on high-impact visual evidence rather than distributing it uniformly.
As the deployment budget decreases from mild to moderate compression (e.g., 192/128/64 tokens), the advantage typically increases because attack effectiveness increasingly depends on corrupting surviving tokens that \attackname targets explicitly.
Under extremely tight budgets (e.g., 32 or 16 tokens), the gain plateaus and may fluctuate on TextVQA and VQA-v2 because robust accuracy approaches the non-visual (blind) reference, leaving little room for further degradation.

% \textbf{Finding 3: \attackname confers consistent advantages across the full compression spectrum, with gains amplifying as the budget tightens.} Even in the non-compressed setting (Full Sequence), \attackname outperforms baselines. We attribute this to EFD's token-reweighting, which concentrates perturbation budget on high-impact visual evidence rather than spreading it uniformly. Crucially, as the token budget decreases (e.g., from 192 to 16), this performance gap \emph{widens}. This indicates that while \attackname acts as a generally stronger encoder-based attack by targeting dominant tokens, its strategic value is maximized in resource-constrained regimes where surviving the bottleneck is the primary determinant of attack success.

% \textbf{Finding 4: Compression is not a reliable defense against attacks.}
% Contrary to the finding that token compression acts as a passive defense by filtering adversarial noise~\cite{gu2025visual}, our results reveal this offers no true immunity.
% While generic baseline attacks indeed lose potency under heavy compression, \attackname successfully targets the selection bottleneck to bypass this ``denoising'' effect.
% For instance, on TextVQA (16 tokens), \attackname drives accuracy down to 15.7\%, indistinguishable from the blind baseline.
% This confirms that compression does not eliminate vulnerability; it merely creates a specific bottleneck that specialized attacks like \attackname can effectively exploit.

\begin{figure}
    \centering
    \includegraphics[width=\linewidth]{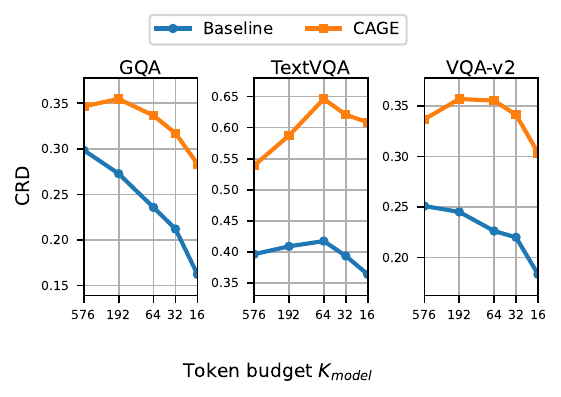}
\caption{\textbf{Conditional robustness degradation ($\mathrm{CRD}$) vs. deployment token budget.} A higher $\mathrm{CRD}$ indicates a larger relative performance drop caused by the attack after accounting for compression-induced clean accuracy degradation. While $\mathrm{CRD}$ under the baseline attack generally decreases as the token budget shrinks, \attackname exhibits non-monotonic behavior.}    
\label{fig:crd}
\end{figure}

\subsection{Analysis and Ablation}
\textbf{Impact of Token Compression on Robustness.}
We investigate the impact of token compression on robustness from two perspectives: absolute robustness and conditional robustness degradation. 
First, under a fixed $\ell_\infty$ perturbation budget, the observed robust accuracy $R$ decreases as the token budget shrinks, reflecting the combined effect of information loss and increased susceptibility to adversarial distortion.
Second, to disentangle robustness degradation from pure information loss, we introduce conditional robustness degradation ($\mathrm{CRD}$), defined as $\mathrm{CRD}=1-\frac{R}{C}$,
where $C$ is the clean accuracy.
$\mathrm{CRD}$ measures the relative fraction of clean performance that is lost under attack, enabling a fairer comparison of attack impact across different compression levels.
A higher $\mathrm{CRD}$ indicates a stronger attack effect after accounting for the clean accuracy drop caused by compression.

\Cref{fig:crd} presents the $\mathrm{CRD}$ trend under the baseline attack and \attackname.
For the baseline attack, $\mathrm{CRD}$ generally decreases as the deployment budget shrinks across all three benchmarks, which can misleadingly suggest that stronger compression improves conditional robustness.
In contrast, \attackname exhibits a non-monotonic pattern: $\mathrm{CRD}$ increases under moderate budgets (e.g., 64--192 tokens) and then partially decreases as the budget shrinks further.
This behavior is consistent with compression-aligned optimization.
Under moderate compression, clean predictions can still be made from the selected survivor tokens, so clean accuracy $C$ drops only mildly.
However, \attackname explicitly targets these same survivors by aligning perturbations with the selection signal, corrupting the evidence that the compressed model relies on.
As a result, robust accuracy $R$ decreases more sharply than $C$, which drives $\mathrm{CRD}$ upward.
Under extremely tight budgets (e.g., 16 tokens), robust accuracy is already severely suppressed, leaving the attack with limited headroom to further reduce $R$, whereas clean accuracy $C$ continues to drop rapidly due to information loss; consequently, $\mathrm{CRD}$ decreases.
Appendix~\ref{appendix:analysis} provides further ablations on components, perturbation budgets, hyperparameters, and scoring metrics.

\section{Possible Defenses}
\label{sec:defense}

\begin{table}[t]
\centering
\caption{Defense performance under different dominant token budgets on VQA-v2. Random denotes a baseline that randomly selects additional tokens to examine whether the robustness gain mainly comes from increased randomness/diversity.}
\label{tab:defense_perf}
\resizebox{\columnwidth}{!}{%
\begin{tabular}{l|cccccc}
\toprule
\textbf{Setting} & \textbf{Full} & \textbf{192} & \textbf{128} & \textbf{64} & \textbf{32} & \textbf{16} \\
\midrule
Clean Acc.              
& 74.5 & 73.4 & 72.1 & 69.3 & 68.4 & 63.2 \\
Robust Acc.     
& 49.4 & 46.5 & 45.1 & 43.0 & 43.1 & 42.7 \\
\midrule
Clean Acc. w/ Random    
& 74.4 & 71.6 & 71.2 & 67.9 & 66.9 & 61.5 \\
Robust Acc. w/ Random   
& 50.6 & 48.4 & 47.7 & 46.5 & 44.9 & 42.8 \\
\midrule
Clean Acc. w/ D1  
& 74.1 & 72.7 & 70.9 & 68.3 & 66.9 & 61.9 \\
Robust Acc. w/ D1     
& 48.2 & 49.9 & 50.0 & 48.4 & 45.3 & 45.5 \\
\midrule
Clean Acc. w/ D2  
& 74.3 & 72.8 & 72.1 & 69.1 & 68.0 & 63.5 \\
Robust Acc. w/ D2    
& 50.7 & 48.6 & 47.6 & 46.6 & 46.1 & 42.8 \\
\bottomrule
\end{tabular}%
}
\end{table}

% \section{Possible Defenses}
% \label{sec:defense}

% Given the vulnerability exposed by \attackname, we explore two potential defense strategies as follows. Detailed description and analysis of the results are provided in the Appendix \ref{appendix:defense}.
We explore two potential defense strategies against \attackname, detailing the results and analysis on existing defenses in Appendix~\ref{appendix:defense}.

\textbf{Selection-based Defenses.}
Since \attackname relies on predicting the deterministic survivor set, we propose two countermeasures:
\textbf{D1 (Robustness-aware Selection)}, which penalizes tokens with unstable attention scores under random noise;
\textbf{D2 (Stochastic Candidate Pool)}, which introduces randomness by sampling survivors from a larger candidate pool.
\Cref{tab:defense_perf} presents the defense performance under different deployment budgets with VisionZIP and VQA-v2. The results show that these defenses are somewhat effective, but the gains remain limited. A key reason is an \emph{informativeness-robustness trade-off}: tokens that are stable under perturbations are often weakly informative. For instance, perturbation-insensitive patches (e.g., backgrounds or low-texture regions) typically lack the semantic density needed for downstream reasoning. As a result, explicitly prioritizing such robust (or random) tokens tends to discard task-critical visual evidence, substantially degrading clean performance and ultimately diminishing the practical utility.

\begin{table}[t]
\centering
\caption{Detection performance using Top-$K$ attention mass as the score on VQA-v2. We report \textbf{Acc.} (Detection accuracy), \textbf{TPR} (True Positive Rate, measuring attack detection success), \textbf{FPR} (False Positive Rate, measuring false alarms on clean inputs), and the \textbf{F1} score.}
\label{tab:det_topk}
\resizebox{0.6\columnwidth}{!}{%
\begin{tabular}{c|cccc}
\toprule
$K$ & Acc. & TPR & FPR & F1 \\
\midrule
1  & 0.900 & 0.850 & 0.050 & 0.895 \\
2  & 0.920 & 0.920 & 0.080 & 0.920 \\
4  & 0.940 & 0.910 & 0.030 & 0.938 \\
% 8  & 0.940 & 0.910 & 0.030 & 0.938 \\
16 & 0.860 & 0.910 & 0.190 & 0.867 \\
% 64 & 0.539 & 1.000 & 0.922 & 0.684 \\
\bottomrule
\end{tabular}}
% \vspace{-3mm}
\end{table}

\textbf{Attention-based Detection.}
We observe that adversarial attacks tend to disperse the attention distribution. Leveraging this finding, we propose a detector based on the Top-$K$ attention mass. Specifically, we compute the cumulative attention assigned to the $K$ highest-attention tokens and use it as a detection score to distinguish adversarial inputs. Results are reported in \Cref{tab:det_topk}, where this simple detector achieves a detection accuracy of 0.94 with $K{=}4$.
However, this detector relies on a threshold tuned on the attacks we evaluate. When the attacker uses a different attack strategy, the attention statistics may shift and the same threshold may no longer work well (see Appendix~\ref{appendix:defense}). Overall, attention-based detection shows promise but necessitates more robust designs that combine multiple statistics.
% This motivates more robust detectors that combine multiple statistics.

% However, the reliance on thresholds calibrated from known attacks limits its generalization to unseen adversarial strategies (in Appendix 
% \ref{appendix:defense}), and motivates more robust, multi-statistic detectors.
% However, the non-negligible false positive rate (14.4\%) remains a hindrance for practical deployment, suggesting that attention dispersion alone provides a strong but not definitive detection signal.

\section{Conclusion}
In this paper, we present the first study on the adversarial robustness of LVLMs under visual token compression. We identify a critical optimization-inference mismatch in existing attacks, which causes overestimating the robustness of compressed LVLMs. To bridge this gap, we propose \attackname, an adversarial attack that aligns adversarial optimization with the compression bottleneck. Extensive experiments demonstrate that \attackname significantly outperforms the SOTA baseline under different deployment token budgets. We further explore potential defenses and find that they offer partial mitigation but remain insufficient. This work serves as a wake-up call for the community to incorporate security evaluations and defenses in the design of efficient LVLMs.

% \newpage
\section*{Acknowledgements}
This work was supported by the Ministry of Science and Technology of the People’s Republic of China (National Key Research and Development Programme, Grant No:  2025YFE0200100), the National Natural Science Foundation of China (Grant No: 62502416), the Research Grants Council (Grant No: 15209922 and 15207725), Hong Kong SAR, China.

\section*{Impact Statement}
This work studies adversarial robustness of large vision-language models (LVLMs) under visual token compression. Since we develop and evaluate attacks, our findings could be misused to degrade deployed systems. We therefore present \attackname as a diagnostic tool: it reveals failure modes introduced by compression and enables more faithful robustness evaluation.
To reduce misuse risk, we do not release any automated pipeline that would directly facilitate real-world abuse. Instead, we focus on controlled experiments, principled analysis, and actionable measurements that support reproducible auditing and defense development. We also explore several potential defenses as initial mitigation steps. Crucially, this study highlights the overlooked security implications of token compression. 
By establishing a rigorous evaluation protocol, we aim to shift the community's focus from purely efficiency-driven designs to mechanisms that are inherently robust against adversarial manipulation.

\bibliography{myref_checked}
\bibliographystyle{icml2026}

%%%%%%%%%%%%%%%%%%%%%%%%%%%%%%%%%%%%%%%%%%%%%%%%%%%%%%%%%%%%%%%%%%%%%%%%%%%%%%%
%%%%%%%%%%%%%%%%%%%%%%%%%%%%%%%%%%%%%%%%%%%%%%%%%%%%%%%%%%%%%%%%%%%%%%%%%%%%%%%
% APPENDIX
%%%%%%%%%%%%%%%%%%%%%%%%%%%%%%%%%%%%%%%%%%%%%%%%%%%%%%%%%%%%%%%%%%%%%%%%%%%%%%%
%%%%%%%%%%%%%%%%%%%%%%%%%%%%%%%%%%%%%%%%%%%%%%%%%%%%%%%%%%%%%%%%%%%%%%%%%%%%%%%
\newpage
\clearpage

\appendix
\onecolumn

\section{Algorithm}
\begin{algorithm}[h]
\caption{\attackname}
\label{alg:caa}
\begin{algorithmic}[1]
\REQUIRE Vision encoder $f_\theta$, clean image $\mathbf{x}$.
\REQUIRE PGD parameters: step size $\alpha$, budget $\epsilon$, iterations $T$.
\REQUIRE \attackname parameters: budget prior $P(K)$, hyperparameter $\lambda$.
\ENSURE Adversarial perturbation $\boldsymbol{\delta}$.

\STATE Initialize $\boldsymbol{\delta}^{(0)} \gets \mathbf{0}$ (or random initialization within $\epsilon$-ball).

\FOR{$t = 0$ to $T-1$}
    \STATE $\mathbf{x}^{(t)}_{adv} \gets \mathbf{x} + \boldsymbol{\delta}^{(t)}$
    \STATE \textcolor{gray}{// Forward pass to get features and scores}
    \STATE Get clean features $\mathbf{z}^{\text{cln}}$, adversarial features $\mathbf{z}^{\text{adv}}$, and attention scores $s(\mathbf{x}^{(t)}_{adv})$ from $f_\theta$.
    
    \STATE \textcolor{gray}{// --- Component I: Expected Feature Disruption (EFD) ---}
    \STATE Compute per-token distortion: $d_i \gets 1 - \cos(\mathbf{z}^{\text{adv}}_i, \mathbf{z}^{\text{cln}}_i)$.
    \STATE Compute current ranks $r_i$ based on scores $s_i$.
    \STATE Compute survival probabilities: $\pi_i \gets \sum_{k} P(K=k) \cdot \mathbb{I}(k > r_i)$.
    \STATE $\mathcal{L}_{\text{EFD}} \gets \frac{\sum \pi_i d_i}{\sum \pi_i}$.
    
    \STATE \textcolor{gray}{// --- Component II: Rank Distortion Alignment (RDA) ---}
    \STATE Compute distortion distribution: $p_d(i) \gets \text{softmax}(d_i)$, with gradients detached.
    \STATE Compute selection distribution: $p_s(i) \gets \text{softmax}(s_i)$.
    % \STATE \textcolor{gray}{// KL divergence with stop-gradient on target}
    \STATE $\mathcal{L}_{\text{RDA}} \gets \sum p_d(i) \cdot \log p_s(i)$.
    
    \STATE \textcolor{gray}{// --- Joint Optimization (Gradient Ascent) ---}
    \STATE $\mathcal{L}_{\text{total}} \gets \mathcal{L}_{\text{EFD}} + \lambda \cdot \mathcal{L}_{\text{RDA}}$.
    \STATE Compute gradient: $\mathbf{g}^{(t)} \gets \nabla_{\boldsymbol{\delta}} \mathcal{L}_{\text{total}}(\mathbf{x}^{(t)}_{adv})$.
    \STATE Update perturbation: $\boldsymbol{\delta}^{(t+1)} \gets \text{Proj}_{\epsilon}\left( \boldsymbol{\delta}^{(t)} + \alpha \cdot \text{sign}(\mathbf{g}^{(t)}) \right)$.
\ENDFOR

\RETURN $\boldsymbol{\delta} \gets \boldsymbol{\delta}^{(T)}$.
\end{algorithmic}
\end{algorithm}

\section{Related Work}
\begin{table}[h]
\centering
\caption{Token compression mechanisms. ``A'' and ``S'' denote attention- and similarity-based metrics, respectively.}
\resizebox{0.6\columnwidth}{!}{
\begin{tabular}{l|c|cc|cc}
\toprule
\multirow{2}{*}{Method} & \multirow{2}{*}{Venue} & \multicolumn{2}{c|}{Outer-LLM} & \multicolumn{2}{c}{Inner-LLM} \\
\cmidrule{3-6}
 & & Prune & Merge & Prune & Merge \\
\midrule
FastV~\cite{chen2024fastv} & ECCV 2024 & \textcolor{red}{\ding{55}} & \textcolor{red}{\ding{55}} & A & \textcolor{red}{\ding{55}} \\
SparseVLM~\cite{zhang2025sparsevlm} & ICML 2025 & \textcolor{red}{\ding{55}} & \textcolor{red}{\ding{55}} & A & S \\
% PDrop~\cite{xing2025pyramiddrop} & arixv 2025 & \textcolor{red}{\ding{55}} & \textcolor{red}{\ding{55}} & A & \textcolor{red}{\ding{55}} \\
DART~\cite{yin2025dart} & arXiv 2025 & \textcolor{red}{\ding{55}} & \textcolor{red}{\ding{55}} & S & \textcolor{red}{\ding{55}} \\
MustDrop~\cite{liu2024mustdrop} & arXiv 2024 & A & S & A & \textcolor{red}{\ding{55}} \\
HoliTom~\cite{shao2025holitom} & arXiv 2025 & \textcolor{red}{\ding{55}} & S & \textcolor{red}{\ding{55}} & S \\
\midrule
% ToMe~\cite{bolya2023tome} & ICLR 2023 & \textcolor{red}{\ding{55}} & S & \textcolor{red}{\ding{55}} & \textcolor{red}{\ding{55}} \\
VisionZip~\cite{visionzip} & CVPR 2025 & A & S & \textcolor{red}{\ding{55}} & \textcolor{red}{\ding{55}} \\
VisPruner~\cite{zhang2025vispruner} & ICCV 2025 & A+S & \textcolor{red}{\ding{55}} & \textcolor{red}{\ding{55}} & \textcolor{red}{\ding{55}} \\
DivPrune~\cite{alvar2025divprune} & CVPR 2025 & S & \textcolor{red}{\ding{55}} & \textcolor{red}{\ding{55}} & \textcolor{red}{\ding{55}} \\
FlowCut~\cite{tongFlowCut2025} & NeurIPS 2025 & A+S & \textcolor{red}{\ding{55}} & \textcolor{red}{\ding{55}} & \textcolor{red}{\ding{55}} \\
PruMerge~\cite{shang2025prumerge} & ICCV 2025 & A & S & \textcolor{red}{\ding{55}} & \textcolor{red}{\ding{55}} \\
G-Prune~\cite{jiang2025gprune} & AAAI 2025 & S & \textcolor{red}{\ding{55}}  & \textcolor{red}{\ding{55}} & \textcolor{red}{\ding{55}} \\
HiRED~\cite{hired2025aaai} & AAAI 2025 & A & \textcolor{red}{\ding{55}}  & \textcolor{red}{\ding{55}} & \textcolor{red}{\ding{55}} \\
\bottomrule
\end{tabular}}
\label{tab:spatial_redundancy}
\end{table}
\textbf{LVLM Efficiency.}
% Vision-language projectors serve as the critical bridge between pre-trained vision encoders and large language models in multimodal architectures. Early approaches primarily employed simple linear projectors that maintain one-to-one correspondence between visual patches and tokens ~\cite{llava,instructblip}. While computationally efficient, these linear projectors often result in an overwhelming number of visual tokens, creating significant computational bottlenecks in downstream processing.
The integration of visual inputs into LLMs leads to long token sequences and substantial computational and memory overhead~\cite{li2024llavanext}. 
Such overhead becomes especially restrictive in latency-sensitive scenarios, including LVLM-driven mobile agents and reasoning-intensive applications, where efficient inference is critical~\cite{yao2025var,yao2026diversity}. 
This has spurred growing interest in visual token compression for LVLMs.
% The integration of visual inputs into LLMs leads to long token sequences and substantial computational and memory overhead~\cite{li2024llavanext}. Such overhead becomes especially restrictive in latency-sensitive scenarios, including LVLM-driven mobile agents, which has spurred growing interest in visual token compression for LVLMs.
Existing token compression approaches can be categorized by \emph{where} token reduction is applied relative to the language model, as shown in \Cref{tab:spatial_redundancy}. 
\ding{172} \textit{Inner-LLM} approaches~\cite{chen2024fastv,zhang2025sparsevlm,yin2025dart,liu2024mustdrop,shao2025holitom}, such as FastV~\cite{chen2024fastv} and SparseVLM~\cite{zhang2025sparsevlm}, integrate token compression into the language model’s transformer layers, reducing the effective number of visual tokens processed during decoding.
\ding{173} \textit{Outer-LLM} approaches perform token selection or aggregation \emph{before} the main language model computation, treating the LLM as a black box.
Representative methods include PruMerge~\cite{shang2025prumerge}, VisionZip~\cite{visionzip}, VisPruner~\cite{zhang2025vispruner}, and FlowCut~\cite{tongFlowCut2025}, which differ mainly in how token importance or redundancy is estimated (e.g., attention-based scoring, similarity-based pruning, or cross-layer information flow), but share a common paradigm of producing a compact visual token sequence prior to LLM inference.
In this work, we focus on \textit{outer-LLM} compression methods, as they are plug-and-play, require no modification to the language model, and are broadly compatible with existing inference and acceleration frameworks.

\textbf{LVLM Robustness.}
The security and privacy problems of deep learning models have received growing attention~\cite{zhang2025merinspector,bai2025toward,bai2025provfl}. In particular, the robustness of LVLMs has emerged as a critical concern as these systems are increasingly deployed in real-world applications, such as medical image analysis \cite{nath2025vila,lin2025healthgpt} and autonomous systems \cite{lubberstedt2025v3lma}. Existing adversarial attacks on LVLMs broadly fall into two optimization paradigms. \ding{172} End-to-end attacks backpropagate through the entire multimodal pipeline to craft adversarial images, but are often computationally expensive due to large models and long contexts \cite{schlarmann2023}. \ding{173} Encoder-based attacks optimize perturbations by targeting only the vision encoder, providing a significantly lighter alternative \cite{vlattack,dongHowRobustGoogles2023,cuiRobustnessLargeMultimodal2024,wangBreakVisualPerception2024,xieChainAttackRobustness2025,zhang2026understanding}. For instance, VT-Attack~\cite{wangBreakVisualPerception2024} perturbs encoded visual tokens to break the vision encoder’s token representations. For prompt-diverse tasks such as VQA, VEAttack~\cite{mei2025veattack} crafts downstream-agnostic examples by minimizing the cosine similarity between clean and perturbed visual token features, and further provides theoretical analysis establishing a lower-bound misalignment guarantee in the LLM-aligned representation space. However, existing encoder-based attacks are typically optimized on full-token models, while token compression reshapes the encoder’s token space, motivating our study of robustness in recently compressed LVLMs.

\section{Detailed Experimantal Setup}
\subsection{Datasets}
\label{appendix:datasets}

We evaluate on diverse datasets that cover complementary domains essential to real-world LVLM applications.

\noindent\textbf{VQA-v2} \cite{goyal2017making} is a large-scale open-ended visual question answering benchmark built on COCO images, where each question is annotated with multiple human answers. It covers a broad spectrum of everyday visual concepts (objects, attributes, actions, counting, and commonsense cues), and is widely used as a representative testbed for general-purpose VQA. In our study, VQA-v2 serves as a \emph{general perception and language grounding} benchmark, reflecting the typical deployment scenario of LVLMs for open-domain visual understanding.

\noindent\textbf{TextVQA}~\cite{singh2019towards} is designed for \emph{scene-text-centric} reasoning: questions often require reading and understanding text embedded in natural images (e.g., signs, product packages, screens, menus), and then integrating it with visual context. This benchmark stresses the model’s OCR-related capability and fine-grained visual grounding under cluttered backgrounds. We include TextVQA because token compression may disproportionately affect small, high-frequency visual cues (such as characters and short words), and thus its robustness behavior can differ substantially from VQA.

\noindent\textbf{GQA}~\cite{hudson2019gqa} focuses on \emph{compositional visual reasoning}, featuring structured questions that frequently require multi-hop reasoning over objects, relations, and attributes (e.g., spatial relations, comparisons, logical conjunctions). Compared to open-ended VQA, GQA emphasizes systematic generalization and relational grounding. We use GQA to test whether compression-aligned attacks remain effective when model decisions rely more heavily on relational evidence and compositional structure rather than isolated salient cues.

\subsection{Compression Mechanism}
\label{appendix:compression_method}

We briefly summarize the visual token compression methods evaluated in this paper.

\noindent\textbf{VisionZip~\cite{visionzip}} adopts a \emph{select-then-merge} pipeline focused on redundancy reduction. It first ranks visual tokens based on importance scores (e.g., attention) and retains the Top-$K$ tokens as survivors. To minimize information loss, the discarded tokens are merged into the semantically similar survivors, thereby reducing the sequence length while preserving relevant visual details. In our experiments, we set the number of merged tokens to 10\% of the whole token budget.

\noindent\textbf{VisPruner~\cite{zhang2025vispruner}} performs \emph{saliency-based pruning}. It utilizes vision-encoder attention maps to estimate token importance, identifying and retaining a compact subset of highly informative tokens. The method explicitly filters out background redundancy to maximize the semantic density of the pruned sequence under a fixed budget.

\noindent\textbf{DivPrune~\cite{alvar2025divprune}} is a \emph{diversity-driven} selection mechanism. Unlike standard methods that prioritize high-activation regions, DivPrune selects a subset of tokens that maximizes feature diversity. This strategy reduces redundancy among retained tokens and ensures broad coverage of distinct image regions, preventing the model from over-focusing on a single salient object.

\noindent\textbf{FlowCut~\cite{tongFlowCut2025}} assesses token redundancy through \emph{attention flow} rather than static importance scores. It analyzes cross-layer token interactions to estimate the information flow, pruning tokens that contribute minimally to the global context. This allows for a more structural reduction of redundancy based on the network's internal information propagation.

\noindent\textbf{PruMerge~\cite{shang2025prumerge}} employs a hybrid \emph{pruning-and-merging} strategy. It first selects a set of representative tokens based on importance and then merges the remaining tokens into these representatives (e.g., via weighted averaging based on similarity). This approach balances the efficiency of pruning with the information preservation of merging.

% \xw{
% TODO (DDL:2026/1/1):

% % 1. Experiments on Qwen.
% 2. Comparison with more baselines.
% % 3. Different perturbation budgets.
% % 4. We use attention to rank the important, further evaluation on $n-norm$.
% 5. Ablation study on temperatures.
% % 6. Ablation study on different designs.
% }

\section{Attack Performance on Other LVLMs}
\label{appendix:qwen}

\subsection{Qwen2.5-VL}
\begin{table*}[t]
\centering
\caption{Attack performance (\%) of various compressed methods on Qwen2.5-VL across different datasets. We report clean accuracy ({Clean}) and robust accuracy under a baseline attack ({Adv (Base)}) and our compression-aligned attack ({Adv (\attackname)}); lower robust accuracy indicates a stronger attack. \textcolor{green!50!black}{Green} annotations indicate the relative decrease of Adv (\attackname) over Adv (Base).}
\label{tab:main_result_Qwen}
\resizebox{\textwidth}{!}{
\begin{tabular}{c|ccl|ccl|ccl}
\toprule
\multirow{2}{*}{\textbf{Compressed Method}} &
\multicolumn{3}{c|}{\textbf{GQA}} &
\multicolumn{3}{c|}{\textbf{TextVQA}} &
\multicolumn{3}{c}{\textbf{VQA-v2}} \\
& Clean & Adv (Base) & Adv (\attackname)
& Clean & Adv (Base) & Adv (\attackname)
& Clean & Adv (Base) & Adv (\attackname) \\
\midrule

% -------------------------------------------------------
\rowcolor{gray!15}\multicolumn{10}{c}{\textit{Upper Bound (144 Tokens)}} \\
None  & 54.4 & 35.5 & 35.2$_{\relDec{0.8}}$
     & 70.3 & 38.0 & 33.9$_{\relDec{10.8}}$
     & 65.5 & 40.3 & 40.2$_{\relDec{0.2}}$ \\
\midrule

% -------------------------------------------------------
\rowcolor{gray!15}\multicolumn{10}{c}{\textit{Retain 72 Tokens ($K_\text{model}=72$)}} \\
VisionZIP {\scriptsize\textsf{(CVPR25)}}    & 54.0 & 33.1 & 33.0 & 66.2 & 29.6 & 24.2 & 57.4 & 35.7 & 28.1 \\
VisPruner {\scriptsize\textsf{(ICCV25)}}    & 52.5 & 36.1 & 36.1 & 62.8 & 29.4 & 26.0 & 54.9 & 36.0 & 33.7 \\
DivPrune  {\scriptsize\textsf{(CVPR25)}}    & 53.3 & 36.4 & 35.1 & 58.4 & 34.0 & 33.8 & 58.4 & 38.9 & 35.1 \\
\rowcolor{green!03}
\textbf{Average} & 53.3 & 35.2 & 34.7$_{\relDec{1.3}}$
& 62.5 & 31.0 & 28.0$_{\relDec{9.7}}$
& 56.9 & 36.9 & 32.3$_{\relDec{12.4}}$ \\
\midrule

% -------------------------------------------------------
\rowcolor{gray!15}\multicolumn{10}{c}{\textit{Retain 36 Tokens ($K_\text{model}=36$)}} \\
VisionZIP {\scriptsize\textsf{(CVPR25)}}    & 47.9 & 32.3 & 30.1 & 56.1 & 19.2 & 13.0 & 53.6 & 28.4 & 27.5 \\
VisPruner {\scriptsize\textsf{(ICCV25)}}    & 47.0 & 34.4 & 33.4 & 42.4 & 19.6 & 17.5 & 50.5 & 33.1 & 31.7 \\
DivPrune  {\scriptsize\textsf{(CVPR25)}}    & 51.1 & 36.5 & 36.0 & 57.7 & 28.6 & 24.1 & 55.3 & 36.2 & 34.9 \\
\rowcolor{green!03}
\textbf{Average} & 48.7 & 34.4 & 33.2$_{\relDec{3.5}}$
                 & 52.1 & 22.5 & 18.2$_{\relDec{19.1}}$
                 & 53.1 & 32.6 & 31.4$_{\relDec{3.7}}$ \\

\midrule

% -------------------------------------------------------
% \rowcolor{gray!15}\multicolumn{10}{c}{\textit{Retain 18 Tokens}} \\
% VisionZIP {\scriptsize\textsf{(CVPR25)}}    & 42.1 & 28.7 & 28.0 & 44.0 & 13.1 &  9.8 & 44.1 & 25.6 & 21.9 \\
% VisPruner {\scriptsize\textsf{(ICCV25)}}    & 43.9 & 30.1 & 27.2 & 29.8 & 10.1 & 11.6 & 29.8 & 27.9 & 30.8 \\
% DivPrune  {\scriptsize\textsf{(CVPR25)}}    & 48.1 & 34.7 & 35.2 & 40.7 & 21.8 & 19.5 & 52.1 & 34.7 & 35.2 \\
% \rowcolor{green!03}
% \textbf{Average}                            & 44.7 & 31.2 & 30.1 \textcolor{green!50!black}{\scriptsize (-3.5\%)} & 38.2 & 15.0 & 13.6 \textcolor{green!50!black}{\scriptsize (-9.3\%)} & 42.0 & 29.4 & 29.3 \textcolor{green!50!black}{\scriptsize (-0.3\%)} \\
% \midrule

\rowcolor{gray!15}\multicolumn{10}{c}{\textit{Blind ($K_\text{model}=0$)}} \\
None  & 28.4& 28.4 & 28.4 & 6.8 & 6.8 & 6.8 & 17.0 & 17.0 & 17.0 \\
\bottomrule
\end{tabular}}
\end{table*}

Table~\ref{tab:main_result_Qwen} reports the attack performance on Qwen2.5-VL under different compression settings. 
To ensure that the token budget is computed accurately, we resize all input images to $336\times336$, yielding a fixed number of visual tokens in Qwen.
We verify attack performance under three representative compression mechanisms, since most existing methods do not release Qwen-compatible implementations and their official codebases are tightly coupled to specific LLaVA.
Overall, our attack (\attackname) consistently achieves lower robust accuracy than the baseline attack across all datasets and token budgets, indicating stronger attack effectiveness. However, compared to the results on LLaVA, the relative improvement brought by \attackname on Qwen2.5-VL is moderate.

This observation can be attributed to the architectural characteristics of Qwen2.5-VL. Since the visual encoder of Qwen natively integrates token merging, the baseline attack is optimized directly within this compressed feature space. This minimizes the optimization-inference mismatch that typically affects models like LLaVA. Consequently, the performance gain of our method appears less pronounced on Qwen2.5-VL than on models using dense patch sets.
% In this sense, the performance gap between the baseline attack and \attackname is naturally narrowed, as some compression-related effects are implicitly handled by the model architecture.
Despite this built-in compression, \attackname still demonstrates clear advantages, especially under more aggressive token budgets. For instance, \attackname achieves average relative decreases reaching up to 19.1\% on TextVQA and $K_{\text{model}}=36$. This indicates that explicitly modeling the interaction between token selection and adversarial perturbations remains beneficial, even for LVLMs that already employ internal token merging.

\subsection{InternVL3-8B}

\begin{table}[t]
\centering
\caption{Attack performance (\%) on InternVL3 and VQA-v2 under different visual token compression settings. We report clean accuracy ({Clean}) and robust accuracy under a baseline attack ({Adv (Base)}) and our compression-aligned attack ({Adv (\attackname)}); lower robust accuracy indicates a stronger attack. \textcolor{green!50!black}{Green} annotations indicate the relative decrease of Adv (\attackname) over Adv (Base).}
\label{tab:main_result_internvl}
\resizebox{0.5\linewidth}{!}{
\begin{tabular}{lccc}
\toprule
\textbf{Compressed Method} & \textbf{Clean} & \textbf{Adv (Base)} & \textbf{Adv (\attackname)} \\
\midrule

\rowcolor{gray!15}
\multicolumn{4}{c}{\textit{Upper Bound (144 Tokens)}} \\
None 
& 71.1 & 43.6 & 40.0$_{\relDec{8.3}}$ \\
\midrule

\rowcolor{gray!15}
\multicolumn{4}{c}{\textit{Retain 72 Tokens ($K_\text{model}=72$)}} \\
VisionZIP {\scriptsize\textsf{(CVPR25)}} 
& 70.8 & 39.9 & 33.4 \\
VisPruner {\scriptsize\textsf{(ICCV25)}} 
& 69.7 & 40.8 & 35.9\\
DivPrune  {\scriptsize\textsf{(CVPR25)}} 
& 69.5 & 41.6 & 38.1 \\
\rowcolor{green!03}
\textbf{Average} 
& 70.0 & 40.8 & 35.8$_{\relDec{12.3}}$ \\
\midrule

\rowcolor{gray!15}
\multicolumn{4}{c}{\textit{Retain 36 Tokens ($K_\text{model}=36$)}} \\
VisionZIP {\scriptsize\textsf{(CVPR25)}} 
& 66.0 & 35.5 & 30.8 \\
VisPruner {\scriptsize\textsf{(ICCV25)}} 
& 66.0 & 36.7 & 33.3\\
DivPrune  {\scriptsize\textsf{(CVPR25)}} 
& 66.3 & 38.3 & 34.4 \\
\rowcolor{green!03}
\textbf{Average} 
& 66.1 & 36.8 & 32.8$_{\relDec{10.9}}$ \\
\bottomrule
\end{tabular}}
\end{table}

Table~\ref{tab:main_result_internvl} further reports the attack performance on InternVL3-8B under different visual token compression settings on VQA-v2.
This experiment verifies whether the effectiveness of \attackname generalizes beyond LLaVA-style architectures and Qwen2.5-VL.
Overall, \attackname consistently achieves lower robust accuracy than the baseline attack across all compression methods and token budgets, demonstrating that compression-aligned perturbation optimization remains effective on InternVL3-8B.
For example, under VisionZIP with $K_{\text{model}}=72$, \attackname reduces the robust accuracy from 39.9\% to 33.4\%, corresponding to a relative decrease of 16.3\%.
Under more aggressive compression with $K_{\text{model}}=36$, \attackname also consistently improves attack effectiveness across VisionZIP, VisPruner, and DivPrune.

These results suggest that the optimization-inference mismatch is not limited to a specific LVLM backbone or compression algorithm.
Although InternVL3-8B adopts a different visual representation and multimodal alignment architecture from LLaVA and Qwen2.5-VL, explicitly modeling the interaction between token compression and adversarial perturbations still provides clear benefits.
This further supports the generality of our compression-aligned attack design.

\section{Additional Ablation and Analysis}
\label{appendix:analysis}

\begin{table}[t]
\centering
\caption{Robust accuracy (\%) comparing the EFD-only variant and the complete \attackname design under different budgets on VQA-v2.}
\label{tab:ablation_efd_caa_transposed}
\begin{tabular}{l|cccccc}
\toprule
\textbf{Method} & \textbf{576 (Full)} & \textbf{192} & \textbf{128} & \textbf{64} & \textbf{32}  & \textbf{16} \\
\midrule
EFD Only & 50.8 & 48.5 & 47.0 & 45.8 & 46.0 &45.7 \\
\attackname      & 49.4 & 46.5 & 45.1 & 43.0 & 43.1 &42.7 \\
\bottomrule
\end{tabular}
\end{table}

\subsection{Component Ablation}
We conduct ablations on \attackname to isolate the impact of each design component on attack performance. Table~\ref{tab:ablation_efd_caa_transposed} reports results on VQA-v2 for LLaVA under VisionZip, comparing an EFD-only variant against the full \attackname across token budgets. Across token budgets, \attackname consistently achieves lower robust accuracy than the EFD-only variant, indicating that adding RDA provides a clear benefit. This result suggests that reweighting distortion alone (EFD) is insufficient: by explicitly steering the selection signal, RDA increases the likelihood that highly perturbed tokens are retained after compression, allowing adversarial evidence to pass through the bottleneck and further reducing robustness.

\subsection{Different Perturbation Budgets}
\begin{table}[t]
\centering
\caption{Robust accuracy (\%) under different token budgets on VQA-v2.}
\label{tab:budget_ablation}
% \resizebox{\columnwidth}{!}{
\begin{tabular}{c|cc|cc|cc}
\toprule
\multirow{2}{*}{$K_\text{model}$} 
& \multicolumn{2}{c|}{\textbf{Budget = 2 / 255}} 
& \multicolumn{2}{c|}{\textbf{Budget = 4 / 255}} 
& \multicolumn{2}{c}{\textbf{Budget = 8 / 255}} \\
& {Baseline} & {\attackname} & {Baseline} & {\attackname} & {Baseline} & {\attackname} \\
\midrule
576 (Full) & 55.8 & 49.4 & 47.1 & 45.5 & 46.2 & 45.2 \\
192        & 55.7 & 46.5 & 46.1 & 44.9 & 42.5 & 42.6 \\
128        & 53.8 & 45.1 & 46.2 & 43.5 & 43.9 & 42.9 \\
64         & 53.5 & 43.0 & 46.1 & 41.2 & 43.2 & 42.2 \\
32         & 51.3 & 43.1 & 45.8 & 42.1 & 43.1 & 42.0 \\
16         & 49.7 & 42.7 & 44.7 & 41.1 & 43.2 & 38.9 \\
\bottomrule
\end{tabular}
\end{table}

We further study how attack effectiveness scales with the perturbation budget.
Table~\ref{tab:budget_ablation} compares the attack performance under different perturbation budgets $\epsilon \in \{2/255,\,4/255,\,8/255\}$. The results show that \attackname consistently outperforms the baseline across all perturbation budgets, with the advantage becoming more pronounced under smaller token budgets.
In particular, in the low-budget regime ($\epsilon{=}2/255$), \attackname provides the largest gains because the perturbation budget is scarce and must be allocated efficiently: the baseline dilutes optimization over tokens that will later be pruned, while \attackname concentrates distortion on the likely-retained survivor tokens, yielding substantially stronger attacks (e.g., 49.7\%$\rightarrow$42.7\% at 16 tokens).
As $\epsilon$ increases, both attacks become stronger and robust accuracy decreases overall, confirming the expected monotonic trend with larger perturbation budgets.
Meanwhile, the relative gap may narrow at higher $\epsilon$ due to diminishing returns: with sufficient budget, the baseline can partially compensate for dilution by injecting stronger distortion, and robust accuracy approaches a performance floor with limited remaining headroom for further degradation.

\subsection{The Impact of $\lambda$}

\begin{table}[t]
\centering
\caption{Impact of hyperparameter $\lambda$ on attack effectiveness under different dominant token budgets on VQA-v2. 
We report robust accuracy (\%), where lower values indicate stronger attacks.}
\label{tab:lambda}
\begin{tabular}{c|cccccc}
\toprule
$K_\text{model}$ & $\lambda=0$ (EFD only) &$\lambda=0.1$ & $\lambda=0.05$ & $\lambda=0.01$ & $\lambda=0.005$ & $\lambda=0.001$ \\
\midrule
576 (Full) & 50.8 &51.8 & 49.5 & 50.5 & \textbf{49.4} & 50.1 \\
192        & 48.5&49.0 & \textbf{46.5} & 46.6 & \textbf{46.5} & 47.7 \\
128        & 47.0& 46.0 & 45.3 & \textbf{44.9} & 45.1 & 45.3 \\
64         & 45.8&45.4 & 45.9 & 43.4 & \textbf{43.0} & 43.8 \\
32         & 46.0&44.8 & 44.6 & \textbf{42.9} & 43.1 & 43.0 \\
16         & 45.7&43.0 & 43.6 & 43.0 & \textbf{42.7} & 43.6 \\
\bottomrule
\end{tabular}
\end{table}

We analyze the sensitivity of \attackname to the weighting coefficient $\lambda$, which balances the two objectives. Results are reported in Table~\ref{tab:lambda}. Overall, moderate values consistently perform best: $\lambda{=}0.005$ yields the lowest accuracy for most budgets (Full/64/16), while $\lambda{=}0.01$ is slightly better in a few cases (128/32) and remains competitive across the board. Importantly, most non-zero $\lambda$ settings are comparable to or better than $\lambda{=}0$, indicating that incorporating RDA is generally beneficial even without delicate tuning. In contrast, overly large $\lambda$ (e.g., 0.1) generally weakens the attack, suggesting that over-emphasizing the auxiliary alignment term can distract optimization from inducing sufficient feature disruption. Meanwhile, extremely small $\lambda$ (e.g., 0.001) also degrades performance, indicating that without enough alignment pressure, adversarial distortion is less likely to survive the compression bottleneck. These results imply that a modest $\lambda$ is crucial to jointly achieve strong feature disruption and effective bottleneck alignment, and we adopt $\lambda{=}0.005$ as a robust default.

% \subsection{The Impact of $\tau$}
% \begin{table}[t]
% \centering
% \caption{Impact of the threshold parameter $\tau$ on attack effectiveness under different dominant token budgets on VQA-v2. 
% Lower accuracy indicates stronger attacks.}
% \label{tab:tau_attack}
% \begin{tabular}{c|cccc}
% \toprule
% \textbf{$K_\text{model}$} & $\boldsymbol{\tau=0.5}$ & $\boldsymbol{\tau=0.2}$ & $\boldsymbol{\tau=0.1}$ & $\boldsymbol{\tau=0.05}$ \\
% \midrule
% 576 (Full) & 50.10 & 50.70 & \textbf{49.20} & 53.20 \\
% 192        & 48.60 & 46.80 & \textbf{45.80} & 50.50 \\
% 128        & 47.70 & \textbf{45.30} & 45.60 & 49.30 \\
% 64         & 45.20 & 44.40 & \textbf{43.90} & 46.90 \\
% 16         & \textbf{42.70} & 44.60 & 43.30 & 45.30 \\
% \bottomrule
% \end{tabular}
% \end{table}

% Table~\ref{tab:tau_attack} analyzes the effect of the threshold parameter $\tau$ from an attack perspective, where lower accuracy indicates stronger attacks.
% We observe that $\tau=0.1$ consistently leads to the lowest accuracy across most dominant token budgets, yielding the strongest overall attack performance.
% In contrast, larger thresholds (e.g., $\tau=0.5$) result in weaker attacks, while overly small values (e.g., $\tau=0.05$) tend to preserve too many stable tokens, significantly reducing attack effectiveness.
% These results suggest that $\tau=0.1$ provides a favorable balance between aggressively perturbing dominant tokens and maintaining sufficient coverage to propagate adversarial effects.

\begin{table}[t]
\centering
\caption{Impact of different score metrics on attack effectiveness.
We report robust accuracy (\%), where lower values indicate stronger attacks.}
\label{tab:score_metric}
\begin{tabular}{c|cc}
\toprule
$K_\text{model}$ & Norm & Attention \\
\midrule
576 (Full) & 65.2 & \textbf{49.4} \\
192        & 65.2 & \textbf{46.5} \\
128        & 62.5 & \textbf{45.1} \\
64         & 60.6 & \textbf{43.0} \\
32        & 56.3 & \textbf{43.1} \\
16         & 51.0 & \textbf{42.7} \\
\bottomrule
\end{tabular}
\end{table}

\subsection{Impact of Token Importance Metric}
We study the impact of different token scoring metrics used in the attack pipeline to determine which signal better guides the adversarial optimization. Specifically, we compare two representative metrics:

\begin{itemize}[leftmargin=*, topsep=0pt, itemsep=0pt]
\item \textbf{$\ell_1$-norm score:} Prior work~\cite{tongFlowCut2025,guo-etal-2024-attention} shows that the $\ell_1$ norm of each token’s value vector is a proxy for information strength, which can measure the token importance. This metric ranks tokens by embedding magnitude, based on the intuition that smaller norms correspond to weaker signals and thus convey less visual information through the Value vectors.

\item \textbf{Attention score:} We rank tokens by their attention-based importance, which reflects each token’s contribution to multimodal reasoning. Compared to norm-based scoring, attention better captures semantic relevance to the textual context. Attention-based importance has been widely adopted in token selection and compression for LVLMs~\cite{visionzip,tongFlowCut2025,hired2025aaai}.
\end{itemize}

As shown in Table~\ref{tab:score_metric}, the attention-based metric consistently leads to significantly lower robust accuracy than the norm-based alternative across all dominant token budgets. This indicates that attention provides a more effective signal for identifying tokens whose perturbation has a larger impact on the model output.
This behavior can be explained by the semantic nature of attention weights in LVLMs. While embedding norms mainly capture low-level activation magnitude, they do not necessarily correlate with a token’s influence on cross-modal alignment or downstream reasoning. In contrast, attention scores explicitly encode how strongly visual tokens interact with language tokens during inference, making them a more reliable proxy for token importance in multimodal decision making.

\subsection{The Impact of budget-prior range $[K_\text{min}, K_\text{max}]$}
\begin{table}[t]
\centering
\caption{Impact of $K_\text{min}$ and $K_\text{max}$ on attack effectiveness under different dominant token budgets on VQA-v2.
We report robust accuracy (\%), where lower values indicate stronger attacks.}
\label{tab:k_min_max}
\begin{tabular}{c|cccc}
\toprule
$K_\text{model}$ & $[16,64]$ & $[16,128]$ & $[16,192]$ & $[16,384]$ \\
\midrule
576 (Full) & 54.2 & 50.2 & 49.4 & \textbf{49.0} \\
192        & 52.4 & 48.1 & 46.5 & \textbf{46.2} \\
128        & 50.0 & 46.9 & \textbf{45.1} & 46.5 \\
64         & 45.6 & 44.3 & \textbf{43.0 }& 45.6 \\
32         & \textbf{42.0} & 45.4 & 43.1 & 45.8 \\
16         & \textbf{42.3} & 43.3 & 42.7 & 44.7 \\
\bottomrule
\end{tabular}
\end{table}

Table~\ref{tab:k_min_max} examines how the budget-prior range $[K_{\min},K_{\max}]$ affects attack strength. Overall, we observe a clear trade-off between \emph{perturbation concentration} and \emph{attack coverage}. With a narrow prior (e.g., $[16,64]$), \attackname concentrates optimization on only the top-ranked tokens, yielding the strongest attacks under aggressive compression where a handful of survivors dominate inference (e.g., $42.0\%$ at budget $=32$). However, this choice is less effective at higher budgets (e.g., $54.2\%$ at Full), since many tail/context tokens remain weakly perturbed, allowing the model to recover using relatively clean complementary evidence. In contrast, widening the prior (e.g., $[16,384]$) spreads perturbations over a larger portion of the token set, improving effectiveness against high-budget models (reducing Full accuracy to $49.0\%$), but introducing budget dilution that weakens the distortion concentrated on the most influential survivors under extreme compression (e.g., $44.7\%$ at budget $=16$). Among the tested settings, $[16,192]$ provides the best overall balance: it preserves strong disruption on dominant survivors in low-budget regimes while maintaining sufficient coverage to degrade performance in high-budget settings, making it a robust default when the deployment budget is unknown. A practical rule is to set $K_{\min}$ to the most aggressive budget you want to be robust against (smallest survivors) and set $K_{\max}$ to the largest budget expected in deployment (largest survivors), while avoiding overly large $K_{\max}$ that causes dilution.

\subsection{Comparing with More Baselines}

\begin{table}[t]
\centering
\caption{Attack performance on more baselines.}
\label{tab:comprasion_baselines}
\begin{tabular}{l|cccccc}
\toprule
\textbf{Method} & \textbf{576 (Full)} & \textbf{192} & \textbf{128} & \textbf{64} & \textbf{32}  & \textbf{16} \\
\midrule
Clean      & 74.5 &73.4&72.1&69.3&68.4&63.2\\
AttackVLM-ii~\cite{zhao2023evaluating} & 62.6 & 62.1 & 62.1 & 60.8 & 57.3 & 52.5 \\
VT-Attack~\cite{wang2024break} & 59.4 & 60.2 & 60.1 & 59.2 & 54.7 & 51.1 \\
VEAttack~\cite{mei2025veattack}   & 55.8& 55.7&53.8&53.5&51.3&49.7 \\
\attackname &49.4 & 46.5& 45.1& 43.0& 43.1& 42.7  \\
\bottomrule
\end{tabular}
\end{table}

In the main paper, we use VEAattack as the primary baseline, since it is the strongest and most relevant visual attack baseline in our setting. To further verify that our conclusions do not depend on a single baseline choice, we additionally compare with two representative attacks, AttackVLM-ii~\cite{zhao2023evaluating} and VT-Attack~\cite{wang2024break}, on VisionZip and VQA-v2.

Table~\ref{tab:comprasion_baselines} reports the adversarial accuracy under different token budgets. Across all compression levels, our method consistently achieves lower adversarial accuracy than all competing baselines, including the full-token setting. This shows that the advantage of \attackname is not limited to a comparison against VEAattack alone, but generalizes to multiple representative LVLM attack baselines. The gains are particularly clear under compressed settings, which is consistent with our main claim that compression-aware attack design is more effective against compressed LVLMs.

\subsection{Attack Performance on Image Captioning}
\begin{table}[t]
\centering
\caption{Attack performance on image captioning task.}
\label{tab:image_captioning}
\begin{tabular}{l|cccccc}
\toprule
\textbf{Method} & \textbf{576 (Full)} & \textbf{192} & \textbf{128} & \textbf{64} & \textbf{32}  & \textbf{16} \\
\midrule
Clean      & 1.2764 & 1.2941 & 1.3014 & 1.1757 & 1.0075 & 0.6941 \\
Baseline   & 0.2202 & 0.2261 & 0.2317 & 0.1998 & 0.1689 & 0.1199 \\
\attackname & 0.1699 & 0.1740 & 0.1285 & 0.1090 & 0.1166 & 0.0850 \\
\bottomrule
\end{tabular}
\end{table}

We further evaluate the attack performance on the image captioning task. Following the same compressed-LVLM setting, we conduct experiments under VisionZip with different token budgets and report results on 100 samples from the COCO dataset. We use CIDEr as the evaluation metric, where lower scores under attack indicate stronger degradation of caption quality.

As shown in Table~\ref{tab:image_captioning}, \attackname consistently achieves lower CIDEr scores than the baseline across all token budgets, demonstrating stronger attack effectiveness not only on VQA but also on generative captioning tasks. The advantage is particularly clear under moderate and strong compression (e.g., 128/64/32 tokens), where compression-aware attack design more effectively disrupts the visual evidence retained by the compressed model. These results suggest that the benefit of \attackname generalizes in different tasks.

\subsection{Runtime Comparison}
\begin{table}[t]
\centering
\small
\caption{Runtime comparison of different attack methods. We report the average runtime per sample, together with the relative runtime normalized to VEAattack. Lower is better.}
\label{tab:runtime}
\begin{tabular}{lcc}
\toprule
\textbf{Method} & \textbf{Avg. time / sample (s)} & \textbf{Relative cost vs. VEAattack} \\
\midrule
VEAattack    & 5.35 & 1.00$\times$ \\
AttackVLM-ii & 5.28 & 0.99$\times$ \\
VT-Attack    & 9.63 & 1.80$\times$ \\
\attackname         & 5.74 & 1.07$\times$ \\
\bottomrule
\end{tabular}
\end{table}
Table~\ref{tab:runtime} compares the runtime of different attack methods. CAGE requires 5.74 seconds per sample, corresponding to only 1.07$\times$ the cost of VEAattack, while remaining substantially more efficient than VT-Attack (9.63 seconds, 1.80$\times$). This suggests that the stronger attack performance of CAGE comes with only marginal additional computational overhead.

\subsection{Black-box Setting }
\begin{table}[t]
\centering
\caption{Attack performance on other encoder.}
\label{tab:transferability}
\begin{tabular}{l|cccccc}
\toprule
\textbf{Method} & \textbf{576 (Full)} & \textbf{192} & \textbf{128} & \textbf{64} & \textbf{32}  & \textbf{16} \\
\midrule
Clean      & 74.5 &73.4&72.1&69.3&68.4&63.2\\
Baseline   & 69.5 & 69.4& 68.7 & 65.5 & 64.4  & 63.0 \\
\attackname & 69.4&69.1&66.9&64.8&63.1&62.2 \\
\bottomrule
\end{tabular}
\end{table}

We further study a fully black-box transfer setting by using a surrogate vision encoder different from that of the target compressed LVLM. Specifically, attacks are optimized on the surrogate encoder {openai/clip-vit-base-patch16}, while evaluation is performed on the target compressed LVLM under VisionZip with different token budgets. For both the baseline and \attackname, we use the same perturbation budget and optimization protocol: $\epsilon=16/255$, step size $\alpha=1/255$, and 100 PGD steps.

The results are shown in Table~\ref{tab:transferability}. Compared with the gray-box setting in the main paper, the additional gain of \attackname\ over the baseline becomes more limited in this fully black-box setting, although \attackname\ still achieves lower adversarial accuracy under most compressed budgets. We believe this behavior is expected. In the fully black-box case, the attacker not only needs the perturbation itself to transfer across encoders, but also needs the \emph{compression-aware signal} to transfer, i.e., the perturbation should remain aligned with the token-selection and retention behavior of the target compressed LVLM. This is substantially more challenging, because such signals are strongly coupled with the target model's visual encoder and compression pipeline.

These results suggest that the main advantage of \attackname\ is strongest in the gray-box regime, where the attacker knows or can closely approximate the victim model's vision encoder and compression mechanism. By contrast, when the visual encoder is entirely unknown, obtaining useful compression-aware signals for the target model becomes difficult, which limits the extra gain over a generic baseline. Importantly, this does not contradict our main claim; rather, it highlights that beyond standard black-box transferability, exploiting the compression-induced bottleneck itself is an additional challenge in the fully black-box setting.

\subsection{Additional Analysis on Budget Uncertainty and Optimizers}
\label{app:budget_optimizer}

We further evaluate whether the improvement of CAGE can be reproduced by simpler budget sampling or stronger optimizers.
First, we compare CAGE with an EOT-style baseline that samples multiple candidate token budgets at each iteration and optimizes the averaged loss over the corresponding retained tokens.
Second, we test CAGE with APGD and MI-FGSM to examine whether the gain depends on the specific optimizer.
The results are reported in Table~\ref{tab:eot_budget} and Table~\ref{tab:optimizer_results}.

\paragraph{Comparison with EOT.}
A natural alternative for handling unknown deployment budgets is to use an EOT-style objective, where the attack samples multiple candidate token budgets during optimization and maximizes the average loss.
As shown in Table~\ref{tab:eot_budget}, CAGE consistently achieves lower robust accuracy than the EOT baseline across all evaluated token budgets.
For example, when $K_{\text{model}}=64$, CAGE reduces the robust accuracy from 49.8\% to 43.0\%.
Similarly, at $K_{\text{model}}=16$, CAGE further reduces robust accuracy from 47.6\% to 42.7\%.
These results suggest that simply averaging over sampled budgets is insufficient.
Unlike EOT, CAGE explicitly models token survival probabilities and further aligns token distortion with the selection ranking, leading to a more structured compression-aware attack objective.

\begin{table}[t]
\centering
\caption{Attack performance (\%) of the EOT-style budget-sampling baseline and CAGE on VQA-v2. Lower robust accuracy indicates a stronger attack.}
\label{tab:eot_budget}
% \resizebox{\columnwidth}{!}{%
\begin{tabular}{lcccccc}
\toprule
\textbf{Method} & \textbf{576} & \textbf{192} & \textbf{128} & \textbf{64} & \textbf{32} & \textbf{16} \\
\midrule
EOT  & 52.2 & 50.9 & 51.1 & 49.8 & 48.6 & 47.6 \\
CAGE & 49.4 & 46.5 & 45.1 & 43.0 & 43.1 & 42.7 \\
\bottomrule
\end{tabular}%

\end{table}

\paragraph{Effect of optimizers.}
We also evaluate whether the advantage of CAGE depends on the use of PGD.
As shown in Table~\ref{tab:optimizer_results}, CAGE consistently outperforms VEAttack under APGD, MI-FGSM, and PGD.
This indicates that the improvement mainly comes from the compression-aligned objective rather than from a particular optimizer.
Interestingly, stronger or more recent optimizers do not necessarily lead to better attack performance in our setting.
This is likely because these optimizers are mostly designed for standard classification objectives, whereas our attack optimizes feature-level disruption in LVLM visual representations.
Therefore, directly transferring these optimizers to the compressed-LVLM setting does not always provide additional benefits.

\begin{table}[t]
\centering
\caption{Attack performance (\%) using different optimizers on VQA-v2. Lower robust accuracy indicates a stronger attack.}
\label{tab:optimizer_results}
% \resizebox{\columnwidth}{!}{%
\begin{tabular}{lcccccc}
\toprule
\textbf{Method} & \textbf{576} & \textbf{192} & \textbf{128} & \textbf{64} & \textbf{32} & \textbf{16} \\
\midrule
VEAttack (APGD)    & 56.0 & 55.9 & 55.0 & 54.9 & 51.0 & 49.6 \\
CAGE (APGD)        & 51.2 & 48.7 & 46.7 & 45.4 & 43.5 & 42.9 \\
\midrule
VEAttack (MI-FGSM) & 55.6 & 56.0 & 54.1 & 53.6 & 51.1 & 49.4 \\
CAGE (MI-FGSM)     & 48.5 & 46.7 & 46.5 & 44.4 & 42.4 & 43.6 \\
\midrule
VEAttack (PGD)     & 55.8 & 55.7 & 53.8 & 53.5 & 51.3 & 49.7 \\
CAGE (PGD)         & 49.4 & 46.5 & 45.1 & 43.0 & 43.1 & 42.7 \\
\bottomrule
\end{tabular}%
\end{table}

\subsection{Additional Attack Scenario: DriveQA}
\label{app:driveqa}

We further evaluate CAGE on DriveQA to examine its effectiveness in a driving-oriented VQA scenario.
This setting is practically relevant because LVLMs are increasingly considered for safety-critical applications such as autonomous driving, where visual token compression may be adopted to reduce inference latency.
As shown in Table~\ref{tab:driveqa_results}, CAGE consistently achieves lower robust accuracy than VEAttack across all token budgets.
For example, under the full-token setting, CAGE reduces robust accuracy from 28.5\% to 21.0\%.
When the model retains only 16 visual tokens, CAGE further reduces robust accuracy from 18.5\% to 14.0\%.

These results show that the optimization-inference mismatch is not limited to general-purpose VQA benchmarks.
Even in a driving-oriented scenario, where the visual questions are more closely related to safety-critical scene understanding, compression-aware perturbation optimization remains more effective than the baseline attack.
This further supports the need to evaluate efficient LVLMs with attacks that explicitly account for the compressed visual-token pathway.

\begin{table}[t]
\centering
\caption{Attack performance (\%) on DriveQA under different visual token budgets. Lower robust accuracy indicates a stronger attack.}
\label{tab:driveqa_results}
% \resizebox{\columnwidth}{!}{%
\begin{tabular}{lcccccc}
\toprule
\textbf{Setting} & \textbf{576} & \textbf{192} & \textbf{128} & \textbf{64} & \textbf{32} & \textbf{16} \\
\midrule
Clean    & 44.5 & 41.5 & 40.0 & 38.0 & 38.5 & 32.0 \\
VEAttack & 28.5 & 24.5 & 22.5 & 18.5 & 18.5 & 18.5 \\
CAGE     & 21.0 & 17.5 & 16.0 & 17.5 & 15.5 & 14.0 \\
\bottomrule
\end{tabular}%
\end{table}

\begin{figure}[t]
    \centering
    \begin{subfigure}[t]{0.4\linewidth}
        \centering
        \includegraphics[width=\linewidth]{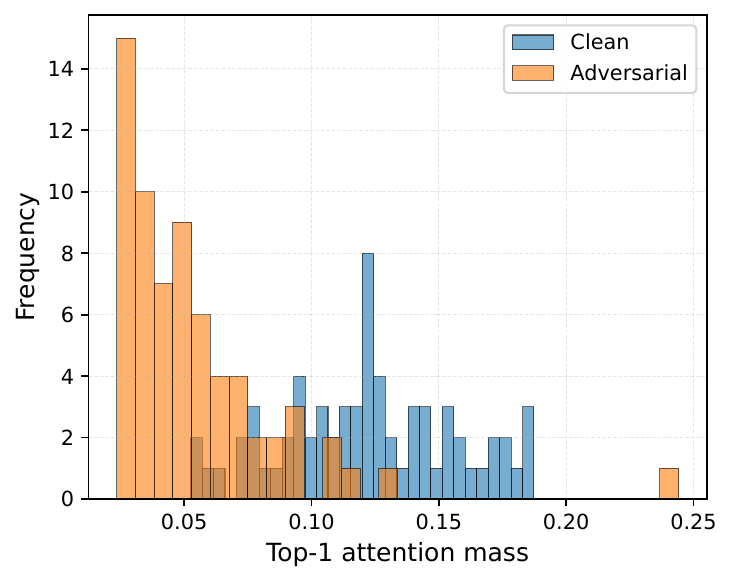}
        \caption{Top-1 attention mass distribution (clean vs. adversarial).}
    \end{subfigure}
    \begin{subfigure}[t]{0.4\linewidth}
        \centering
        \includegraphics[width=\linewidth]{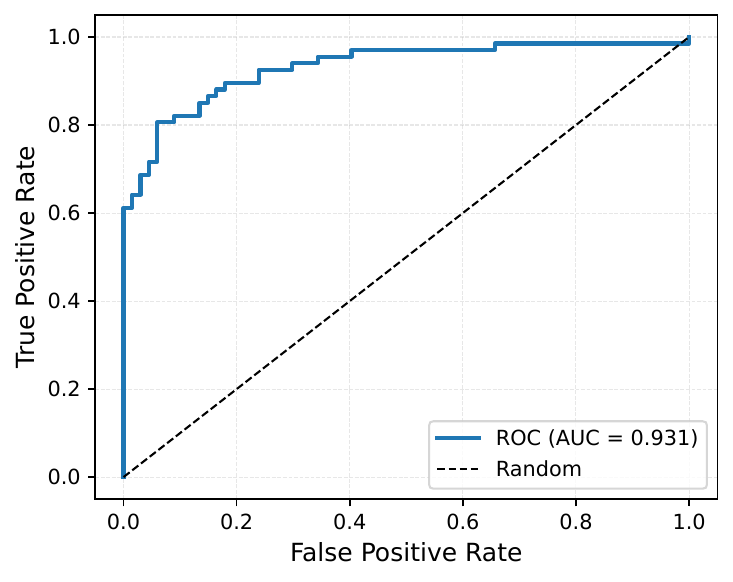}
        \caption{ROC curve using top-1 attention mass as the detection score.}
    \end{subfigure}
    \begin{subfigure}[t]{0.4\linewidth}
        \centering
        \includegraphics[width=\linewidth]{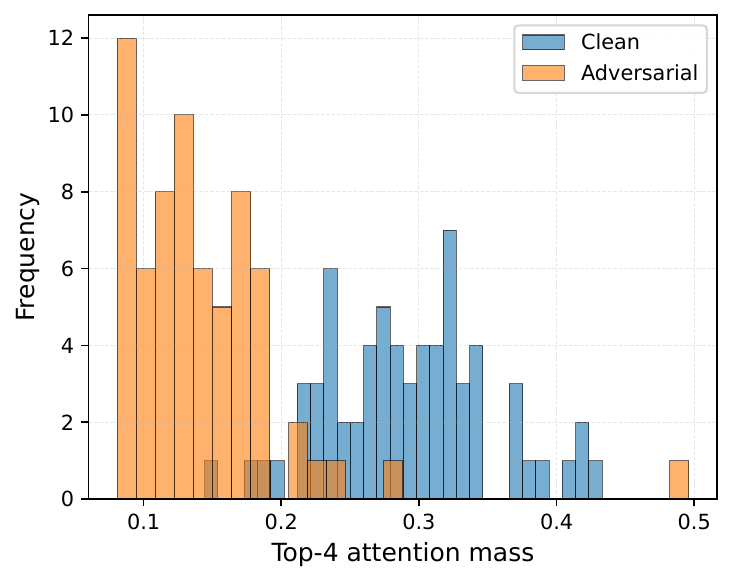}
        \caption{Top-4 attention mass distribution (clean vs. adversarial).}
    \end{subfigure}
    \begin{subfigure}[t]{0.4\linewidth}
        \centering
        \includegraphics[width=\linewidth]{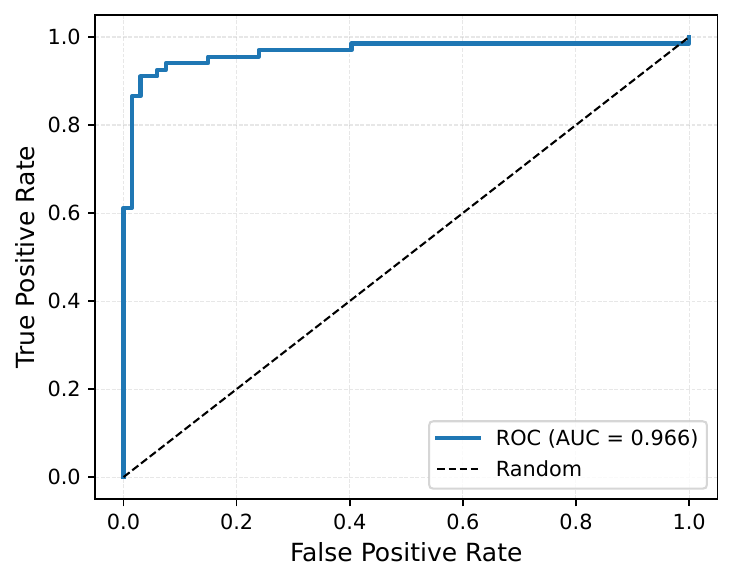}
        \caption{ROC curve using top-4 attention mass as the detection score.}
    \end{subfigure}
    \begin{subfigure}[t]{0.4\linewidth}
        \centering
        \includegraphics[width=\linewidth]{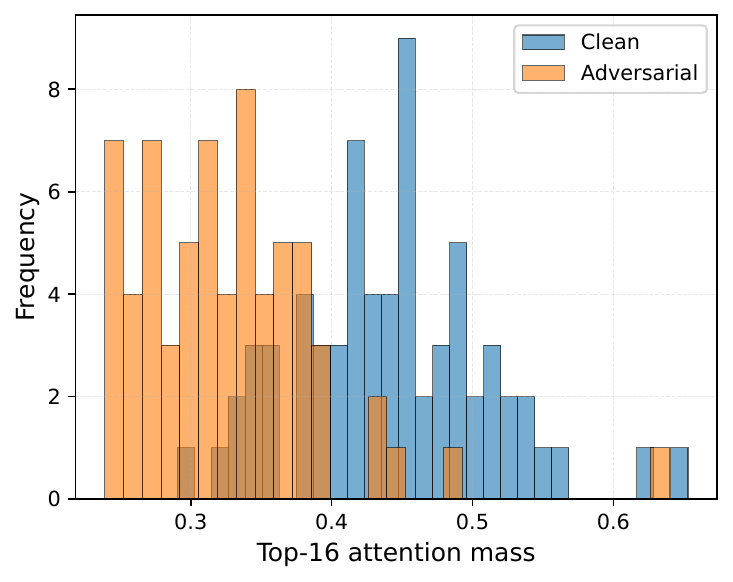}
        \caption{Top-16 attention mass distribution (clean vs. adversarial).}
    \end{subfigure}
    \begin{subfigure}[t]{0.4\linewidth}
        \centering
        \includegraphics[width=\linewidth]{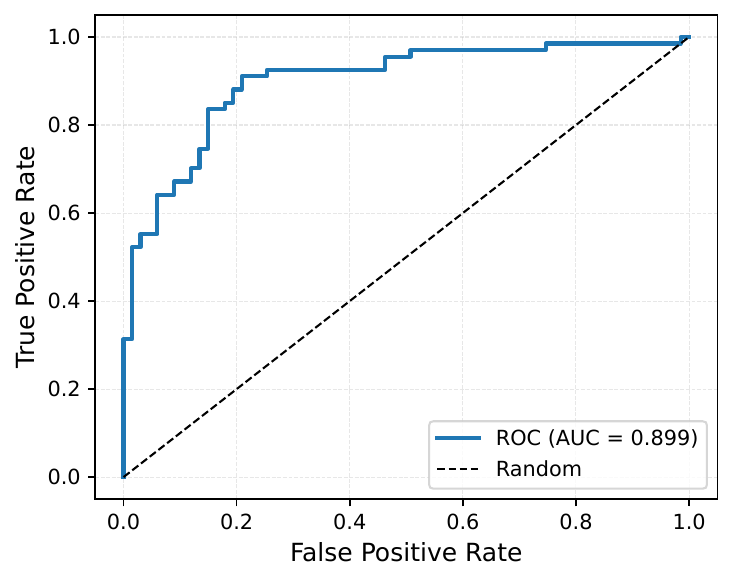}
        \caption{ROC curve using top-16 attention mass as the detection score.}
    \end{subfigure}
    \caption{Attention-based adversarial detection via the top-k CLS-to-token attention mass.}
    \label{fig:attn_detect}
\end{figure}

\section{Possible Defenses}
\label{appendix:defense}

\subsection{Exisiting Defenses}
\begin{table}[t]
\centering
\caption{Defense performance of RobustCLIP.}
\label{tab:robustclip_defense}
% \resizebox{0.85\linewidth}{!}{
\begin{tabular}{lccccc}
\toprule
\textbf{Method} & \textbf{256 (Full)} & \textbf{128} & \textbf{64} & \textbf{32} & \textbf{16} \\
\midrule
Clean Accuracy & 72.0 & 72.2 & 70.4 & 67.5 & 64.5 \\
Adv Accuracy & 45.1 & 44.8 & 43.3 & 42.9 & 41.4 \\
\midrule
Clean Accuracy w/ RobustCLIP & 70.4 & 69.2 & 69.9 & 66.8 & 64.6 \\
Adv Accuracy w/ RobustCLIP & 64.9 & 64.1 & 62.4 & 59.0 & 58.4 \\
\bottomrule
\end{tabular}
\end{table}

\begin{table}[t]
\centering
\caption{Defense performance of DPS \cite{DPS}.}
\label{tab:defense_perf_DPS}
% \resizebox{\columnwidth}{!}{%
\begin{tabular}{l|cccccc}
\toprule
\textbf{Setting} & \textbf{Full} & \textbf{192} & \textbf{128} & \textbf{64} & \textbf{32}  &\textbf{16} \\
\midrule
Clean Acc.              & 74.5 & 73.4 & 72.1 & 69.3 & 68.4& 63.2 \\
Robust Acc.     & 49.4 & 46.5 & 45.1 & 43.0 & 43.1 &42.7 \\
Robust Acc. w/ DPS     & 64.4& 64.6& 63.1& 61.7 &57.6& 49.1\\ 
\bottomrule
\end{tabular}%
\end{table}

Existing defenses against visual adversarial attacks on LVLMs can be broadly grouped into two categories. The first category improves robustness by strengthening the vision encoder itself, typically through adversarial training or robust fine-tuning~\cite{RobustCLIP,dong2025improving}. 
A representative example is RobustCLIP~\cite{RobustCLIP}, which adversarially fine-tunes the CLIP vision encoder to enhance the robustness of downstream systems built upon CLIP features. To examine its effect under visual token compression, we conduct VQA experiments on LLaVA-v1.5-7B while controlling the vision encoder and input resolution. Specifically, we evaluate both the standard and robust victims at $224\times224$, keep the multimodal projector and LLM unchanged, and only replace the vision tower with either the standard CLIP ViT-L/14 encoder or the training-based robust CLIP encoder FARE, using consistent preprocessing in both settings.
As shown in \Cref{tab:robustclip_defense}, RobustCLIP indeed brings substantial robustness gains across all token budgets. 
For example, under the full-token setting, the adversarial accuracy increases from 45.1\% to 64.9\%, while the clean accuracy only slightly decreases from 72.0\% to 70.4\%. 
This suggests that robust visual representation learning is an effective way to improve the adversarial robustness of LVLMs. 
However, dopting RobustCLIP requires replacing and adversarially fine-tuning the visual encoder, which introduces substantial training cost and changes the victim architecture itself. 

The second category consists of training-free, inference-time defenses that operate in a black-box manner. A recent representative method is defense through partial-perception supervision (DPS) \cite{DPS}, which uses responses from partial-image perception to supervise the model’s answer on the original image. DPS is appealing because it does not require retraining and is applicable to black-box LVLMs. As shown  in \Cref{tab:defense_perf_DPS}, DPS indeed yields non-trivial robustness gains, suggesting that response correction from partial views can mitigate a portion of attacked inputs. However, its inference cost is substantially higher, as it requires multiple partial-perception queries together with an additional supervision step during decoding. This overhead makes DPS less suitable for the efficiency-sensitive compressed-LVLM setting considered here. Since token compression is introduced precisely to reduce inference cost, a defense that incurs repeated queries and extra generation steps becomes less practical in this regime. Therefore, while DPS is a reasonable generic inference-time defense, its computational overhead weakens its suitability as a defense baseline for compressed LVLMs.

\subsection{Selection-based Selection}

Token compression introduces a new attack surface: the model’s prediction becomes disproportionately dependent on a small set of surviving tokens. Compression-aligned adversaries can therefore amplify their impact by shaping which tokens pass through the bottleneck and concentrating distortion on these high-impact survivors. This motivates selection-based defenses that intervene at the token selection stage rather than post hoc denoising. In this section, we introduce two \emph{selection-based} defenses that mitigate such predictability from two complementary angles: (i) selecting \emph{robust} tokens whose importance is stable under small perturbations, and (ii) randomizing the survivor set via a \emph{stochastic candidate pool}.

\textbf{D1: Robustness-Aware Selection.}
% \label{sec:defense_selection_robust}
Let $a(x)\in\mathbb{R}^{N-1}$ denote the CLS-to-token attention scores at the penultimate attention layer of the vision encoder for an input image $x$, excluding the CLS token, where $N$ is the number of visual tokens including CLS.
We estimate attention stability using $M$ noisy views $\{x+\delta_m\}_{m=1}^{M}$, where $\delta_m$ is small i.i.d.\ Gaussian noise with pixel-level scale $\eta$, and the perturbed image is clipped to the valid pixel range.
We define a robustness-aware importance score for token $i\in\{1,\dots,N-1\}$ as
\begin{equation}
s_i(x)= \mathbb{E}_{m}\!\left[a_i(x+\delta_m)\right]
-\beta \cdot \mathrm{Std}_{m}\!\left[a_i(x+\delta_m)\right],
\label{eq:robust_score}
\end{equation}
where $\mathbb{E}_m[\cdot]$ and $\mathrm{Std}_m[\cdot]$ are computed over $m=1,\dots,M$, and $\beta$ controls the penalty on attention instability.
Tokens are ranked by $s_i(x)$ and the top-$K_{\text{dom}}$ are selected as dominant survivors.
In our experiments, we use $M=4$, $\eta=1/255$, and $\beta=2.0$.

\textbf{D2: Stochastic Candidate Pool.}
% \label{sec:defense_stochastic}
To reduce the predictability of deterministic Top-$K$ selection, we introduce a simple randomized bottleneck. 
Given an importance score $\tilde{s}\in\mathbb{R}^{N-1}$, we first take a slightly larger candidate set by selecting the top-$(K+\Delta)$ tokens, then uniformly sample $K$ survivors from it:
\[
\mathcal{C}=\text{Top-}(K+\Delta)(\tilde{s}),\qquad 
\mathcal{S}\sim \mathrm{Unif}\big(\mathcal{C}, K\big),
\]
where $\mathrm{Unif}(\mathcal{C},K)$ denotes sampling $K$ elements from $\mathcal{C}$ without replacement.
When $\Delta=0$, this reduces to deterministic Top-$K$; larger $\Delta$ increases randomness and makes the survivor set harder to predict across runs. Generally, we set $\Delta = K$. In practice, if $K+\Delta \ge N-1$, the candidate pool covers all non-CLS tokens and the procedure degenerates to uniformly sampling $K$ tokens from the full set (i.e., purely random selection). Formally, we set the effective pool size as $K+\Delta \leftarrow \min(K+\Delta,\, N-1)$.

\textbf{Experimental Results.}
As shown in Table~\ref{tab:defense_perf}, D1 improves robust accuracy under moderate compression budgets but tends to slightly reduce clean accuracy, and it can even hurt robustness in the full-token setting.
In contrast, D2 largely preserves clean accuracy and provides modest robustness gains at moderate budgets, while becoming ineffective (or even harmful) under the most extreme low-token regime.
Overall, although both defenses offer partial improvements, the adversarial robustness remains limited, likely due to an inherent informativeness–robustness trade-off: tokens that are most stable under perturbations are often less semantically informative for downstream tasks.

% \begin{table}[t]
% \centering
% \caption{Detection performance using Top-$K$ attention mass as the score.}
% \label{tab:det_topk}
% % \resizebox{\columnwidth}{!}{%
% \begin{tabular}{c|cccc}
% \toprule
% $k$ & Acc. & TPR & FPR & F1 \\
% \midrule
% 1  & 0.862 & 0.778 & 0.054 & 0.850 \\
% 16 & 0.907 & 0.958 & 0.144 & 0.912 \\
% 64 & 0.539 & 1.000 & 0.922 & 0.684 \\
% \bottomrule
% \end{tabular}
% \end{table}

\subsection{Attention-based Adversarial Detection}
\label{sec:defense_detection}

\begin{table}[t]
\centering
\caption{Detection performance using Top-$K$ attention mass as the score under cross-attack evaluation settings (threshold calibrated on VEAttack). We report \textbf{Acc.} (Detection accuracy), \textbf{TPR} (True Positive Rate, measuring attack detection success), \textbf{FPR} (False Positive Rate, measuring false alarms on clean inputs), and the \textbf{F1} score.}
\label{tab:det_topk_other_attack}
% \resizebox{\columnwidth}{!}{%
\begin{tabular}{c|cccc}
\toprule
$K$ & Acc. & TPR & FPR & F1 \\
\midrule
1  & 0.877 & 0.892 & 0.137 & 0.879 \\
2  & 0.907& 0.898 & 0.083 & 0.906 \\
4  & 0.925 & 0.934 & 0.083 & 0.925 \\
% 8  & 0.940 & 0.910 & 0.030 & 0.938 \\
16 & 0.772 & 0.580 & 0.035 & 0.718 \\
% 64 & 0.539 & 1.000 & 0.922 & 0.684 \\
\bottomrule
\end{tabular}
\end{table}

In addition to selection-based defenses, we also investigate whether adversarial inputs can be detected from their attention signatures. Our key observation is that adversarial perturbations tend to \emph{reshape} the vision encoder’s CLS-to-token attention distribution, making it less concentrated on a small subset of salient tokens.
Consequently, the cumulative attention mass carried by the most-attended tokens becomes a discriminative signal for separating clean and adversarial inputs.

Let $a(x)\in\mathbb{R}^{N-1}$ denote the CLS-to-token attention scores.
We normalize it into a probability mass function $\hat{a}(x)$ by $\hat{a}_i(x)=a_i(x)/\sum_j a_j(x)$, and then define the \emph{Top-$K$ attention mass} as:
\begin{equation}
m_k(x)=\sum_{i\in \mathrm{Top}\text{-}K(\hat{a}(x))}\hat{a}_i(x),
\label{eq:topk_mass}
\end{equation}
where $\mathrm{Top}\text{-}K(\hat{a}(x))$ returns the indices of the $K$ largest entries of $\hat{a}(x)$.
Intuitively, $m_K(x)$ measures how much attention is concentrated in the most salient $K$ tokens: larger values indicate a more peaked attention pattern.

Following our implementation, we use $\ell(x)=-m_K(x)$ as the adversarial score and classify an input as adversarial if $\ell(x)\ge \tau$. 
We set $\tau$ using a training split: we randomly sample $40\%$ of the available (clean/adversarial) examples as the training set and choose $\tau$ to maximize the Youden index (TPR$-$FPR) on its ROC curve; the remaining $60\%$ of the data is used for evaluation.

Figure~\ref{fig:attn_detect} shows that the Top-$K$ mass yields a clear separation between clean and adversarial examples, and the ROC curve indicates strong discriminative power (AUC $\approx 0.966$ on the training split with $K=4$).
Importantly, this detector is lightweight: it only requires reading attention maps from a single forward pass and computing a scalar statistic.

\textbf{Experimental Results.}
\Cref{fig:attn_detect} visualizes the distribution of Top-$K$ attention mass and the corresponding ROC curve, showing clear separability between clean and adversarial inputs.
On the training split used for threshold selection, the ROC curve yields a strong AUC (e.g., $\mathrm{AUC}\approx 0.96$), indicating that attention concentration provides a discriminative signal for detection.
Table~\ref{tab:det_topk} reports detection performance using the Top-$K$ attention mass as the scoring statistic. 
Overall, we observe that a \emph{small-to-moderate} $K$ provides the most discriminative signal, while overly large $K$ can degrade robustness by blurring the contrast between clean and adversarial attention patterns. 
Specifically, $K=4$ achieves the best overall balance (Acc.=0.940, F1=0.938) with a low false-alarm rate (FPR=0.030), indicating that adversarial examples primarily reduce the concentration of attention on the most salient few tokens. 
In contrast, increasing $K$ to 16 substantially raises false positives (FPR=0.190) without improving detection sensitivity (TPR remains 0.910), which lowers both Acc. and F1. 
This suggests that aggregating attention mass over too many tokens dilutes the “peakiness” cue: clean inputs may still allocate non-trivial attention beyond the top few tokens, making the Top-$K$ mass less separable. 
Therefore, we use a moderate $K$ (e.g., $K=4$) as the default setting in our detector. 

\noindent\textbf{Generalization to Unseen Attacks.}
To evaluate the practical robustness of the proposed detector, we conduct a cross-attack evaluation where the detection threshold $\tau$ is calibrated using adversarial examples from VEAttack, rather than the target attack itself.
As shown in Table~\ref{tab:det_topk_other_attack}, this calibration mismatch leads to a notable degradation in performance.
Most critically, we observe a sharp increase in the FPR. For instance, at the previously optimal setting of $K=4$, the FPR rises to $8.38\%$, nearly tripling the rate observed in the self-calibrated setting ($\sim3\%$).
This indicates that the ``attention dispersion'' boundary learned from VEAttack is ill-suited for the target distribution, causing the detector to aggressively misclassify clean inputs as adversarial.
These findings highlight a fundamental limitation: reliance on a single scalar statistic renders the detector brittle to shifts in attack strategies, necessitating the development of more comprehensive, multi-feature detection frameworks.

\section{Limitations and Future Work}
\label{sec:limitation}

While our work provides the first comprehensive study on the adversarial robustness of LVLMs under visual token compression and proposes the effective \attackname, several limitations remain that pave the way for future research.

\noindent\textbf{Scope of Compression Mechanisms.}
Our study primarily focuses on \emph{training-free, plug-and-play} visual token compression methods (e.g., VisionZip, VisPruner) applied to off-the-shelf LVLMs.
While these represent a widely used paradigm for efficient deployment, we have not explored inner-LLM or training-based compression methods (e.g., FastV \cite{chen2024fastv}, Honeybee \cite{honeybee} or MobileVLM \cite{chu2024mobilevlm}) .
Since these methods modify the compression mechanism and the resulting token dynamics in fundamentally different ways, their robustness characteristics may differ from the settings studied here, motivating a systematic extension of our analysis to these alternative compression families.

\noindent\textbf{Generalization to videos, multi-image inputs, and agentic settings.}
We focus on image-based LVLM tasks and a representative set of benchmarks.
However, many real-world applications may go beyond single images: models often process videos or multiple images, and are increasingly used in agentic workflows that iterate over observe--reason--act steps.
These settings introduce additional challenges, e.g., temporal consistency across frames, cross-image evidence aggregation, and error accumulation over multiple steps, which may interact with token compression in ways not captured by our current evaluation.
A natural next step is to study whether compression-aligned attacks remain effective under such inputs and long-horizon decision loops, and to characterize how the compression bottleneck shapes robustness when visual evidence evolves over time.

\noindent\textbf{More robust defenses.}
Our defense study indicates that lightweight detectors based on a single attention statistic and a calibrated threshold can work well when the calibration attack matches the test-time threat model, but their performance drops under attack distribution shifts (e.g., trained on VEAttack, tested on other strategies), revealing limited generalization.
This suggests that current defenses for compressed LVLMs are still fragile and can overfit to specific attack signatures.
Future work should develop more robust defenses, e.g., combining multi-layer/multi-statistic signals and incorporating the compression bottleneck into robustness objectives.

In summary, we hope this work serves as a stepping stone for designing next-generation LVLMs that are both efficient and secure by design.

\end{document}